%% file: main.tex
\documentclass[aps, twocolumn, superscriptaddress, nofootinbib, nobibnotes,prl]{revtex4-2}

\usepackage[english]{babel}

\usepackage{amsmath,amssymb,bm}
\usepackage{graphicx}
\usepackage{hyperref}
\usepackage{natbib}
\usepackage{braket}
\usepackage[normalem]{ulem}

\usepackage[utf8]{inputenc}

\usepackage{dsfont}
\usepackage{lmodern}
\usepackage[dvipsnames]{xcolor}

\usepackage{mathtools}
\usepackage{xcolor}

\DeclareMathOperator{\Z}{\mathds{Z}}

\newcommand{\hc}{\mathrm{h.c.}}
\newcommand{\nmax}{{N_{\rm max}}}
\DeclareMathOperator{\unit}{\mathds{1}}

\newcommand{\getJuelichAffiliation}{\affiliation{Institute of Quantum Control (PGI-8), Forschungszentrum Jülich, D-52425 Jülich, Germany}}
\newcommand{\getRegensburgAffiliation}{\affiliation{Faculty of Informatics and Data Science, University of Regensburg, D-93040 Regensburg, Germany}}
\newcommand{\getKoelnAffiliation}{\affiliation{Institute for Theoretical Physics, University of Cologne, D-50937 Köln, Germany}}
\newcommand{\getAugsburgAffiliation}{\affiliation{Theoretical Physics III, Center for Electronic Correlations and Magnetism, Institute of Physics, University of Augsburg, D-86135 Augsburg, Germany}}
\newcommand{\getCAAPSAffiliation}{\affiliation{Centre for Advanced Analytics and Predictive Sciences (CAAPS), University of Augsburg, Universitätsstr. 12a, 86159 Augsburg, Germany}}

\begin{document}

\title{Roughening dynamics of interfaces in the two-dimensional quantum Ising model}

\date{\today}

\author{Wladislaw Krinitsin}
~\thanks{These authors contributed equally to this work.} 
\getJuelichAffiliation\getRegensburgAffiliation
\author{Niklas Tausendpfund}
\thanks{These authors contributed equally to this work.} 
\getJuelichAffiliation\getKoelnAffiliation 
\author{Matteo Rizzi}\getJuelichAffiliation\getKoelnAffiliation
\author{Markus Heyl}\getAugsburgAffiliation\getCAAPSAffiliation
\author{Markus Schmitt}
\email{markus.schmitt@ur.de}
\getJuelichAffiliation\getRegensburgAffiliation

\begin{abstract}
The properties of interfaces are key to understand the physics of matter. However, the study of quantum interface dynamics has remained an outstanding challenge. Here, we use large-scale Tree Tensor Network simulations to identify the dynamical signature of an interface roughening transition within the ferromagnetic phase of the 2D quantum Ising model. For initial domain wall profiles we find extended prethermal plateaus for smooth interfaces, whereas above the roughening transition the domain wall decays quickly. Our results can be readily explored experimentally in Rydberg atomic systems.
\end{abstract}

\maketitle

\paragraph{Introduction.}

The static and dynamical properties of interfaces are fundamental to understand the physics and to engineer the functionality of materials.
Interfaces can even undergo their own unique phase transitions independent of the bulk matter.
At a roughening transition, first identified in models for surface growth and classical magnetism \cite{Burton1951,Dobrushin_1973,Weeks1973}, the nature of interfaces changes qualitatively from being smooth to rough.
Intuitively these different phases can be characterized by their interface fluctuations, with small and bounded fluctuations in the smooth phase and large, and unbound ones in the rough phase.
The theoretical analysis of effective surface-on-surface (SOS) models revealed the Berezinski-Kosterlitz-Thouless (BKT) nature of the transition \cite{Chui_76, Beijeren_1977}, which was confirmed in numerical simulations \cite{Hasenbusch_1996}.
For experimental studies, the interface between liquid and crystallized Helium-4 is a particularly suited model system \cite{Balibar_2005}.
At the involved low temperatures, however, quantum effects enter. 
It was found that quantum fluctuations can cause a roughening transition  at vanishing temperature $T=0$ in two-dimensional systems, but not in three dimensions \cite{Fisher1983,Fradkin1983}.
Interface properties are relevant also in a broader scientific context beyond condensed matter.
For instance, in high-energy physics, the flux tube connecting quarks realizes an interface that can undergo roughening~\cite{Luescher_1981}.
Signatures of which were recently observed via digital quantum simulation of a lattice gauge theory \cite{Cochran2024}.
The exploration of such interface roughening in quantum matter \emph{away from equilibrium} has, however, remained an outstanding challenge to date.

In this work we study the non-equilibrium signatures of the roughening transition in the quantum Ising model.
Based on large-scale Tree Tensor Network simulations for real-time evolution \cite{Kloss_2020,Pavesic_2024,krinitsin2025_quantumising}, we find that the underlying quantum roughening transition is reflected in a qualitative change of dynamical behavior of domain wall initial conditions upon tuning the transverse field strength.
In particular, we identify the independent equilibration of the interface in the smooth interface regime as an alternative cause of prethermal plateaus, which is distinct from the known mechanisms relying on approximate conservation laws or Hilbert space fragmentation \cite{Kehrein2008,Moudgalya2022}.
The phenomenology can be readily explored in experiments with Rydberg atomic systems~\cite{Greiner_2002,Georgescu_2014,Kinoshita_2006,Martinez_2016,Jae-yoon_2016,Gross_2017,Jurcevic_2017,Bernien_2017,Zhang_2017,Gaerttner_2017,Choi_2017,Levine_2018,Hild_2014,Barredo_2018,Manovitz_2024}.

\paragraph{Model.}

For the central objective to study the dynamics of interfaces we consider the paradigmatic transverse-field Ising model (TFIM) on a square lattice, given by
\begin{equation}
\mathrm{H} = - J \cdot \sum_{\langle i,j \rangle} \sigma^x_i \sigma^x_j - g \cdot \sum_i \sigma^z_i\,, \label{TFIM}
\end{equation}
where the first sum runs over all neighboring spin pairs. 
The TFIM exhibits a ferromagnetic phase, that extends to non-zero temperatures, with a quantum phase transition at $g_c/J \approx 3.04$ and a thermal transition temperature of $T_c/J \approx 2.27$ at $g=0$~\cite{Bloete_2002}.
Importantly, the 2D quantum Ising model inherits a second quantum phase transition point of BKT type within the ferromagnetic phase, which is associated with a transition from smooth to rough interfaces~\cite{Fradkin1983}. A suited order parameter will be defined at a later point of this work, providing a quantitative description of the transition.
In the quantum domain, when considering the limit $J \gg g$, the TFIM can be perturbatively linked to the PXP model, which shows strong dynamical constrains and Hilbert space fragmentation, leading to a slow relaxation of various domain wall initial conditions~\cite{Balducci_2022,Balducci_2023,krinitsin2025_quantumising, Pavesic_2024}.

\begin{figure}[t!]
    \begin{center}
    \includegraphics[width=\linewidth]{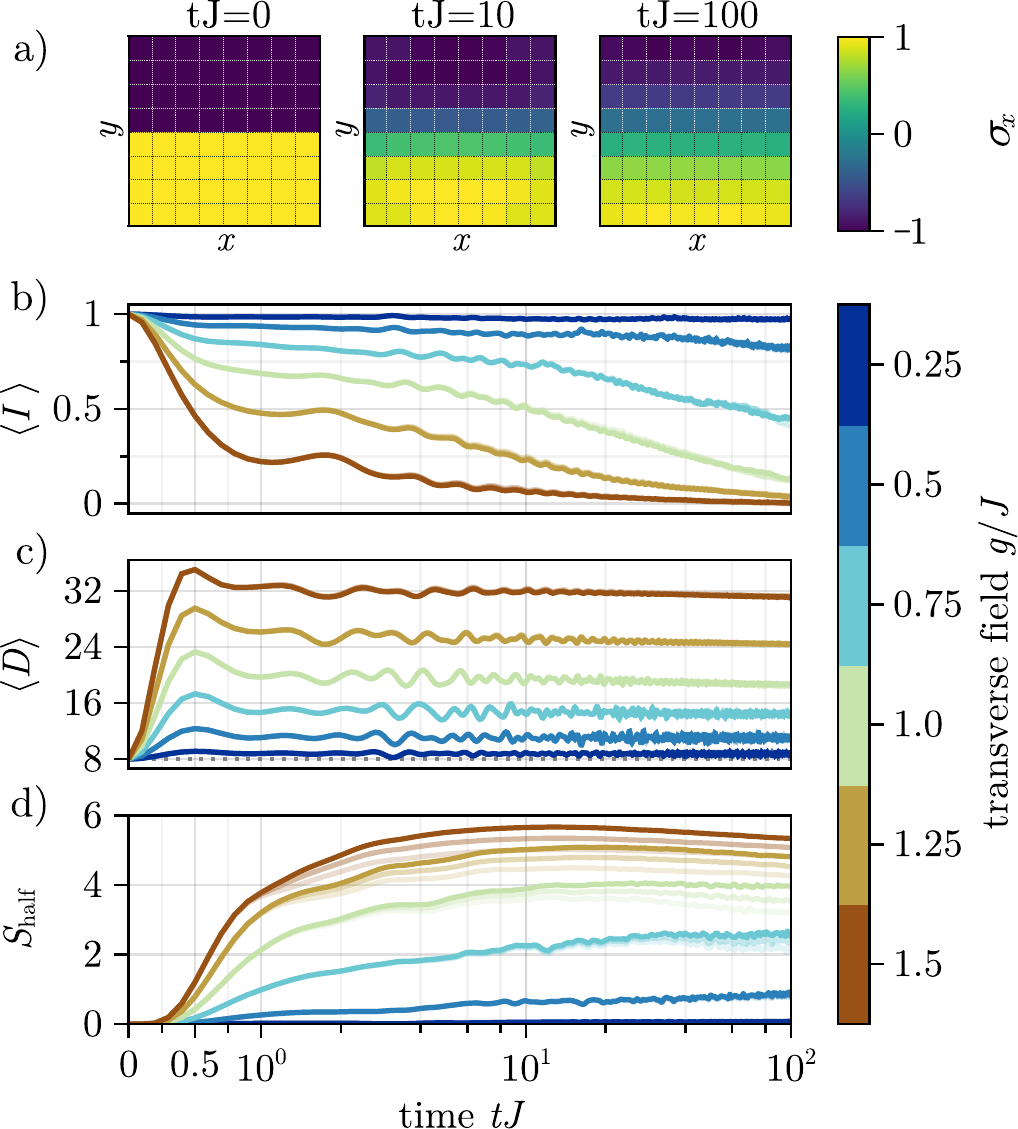}
    \end{center}
    \caption{Time evolution of a flat interface on an 8$\times$8 lattice with open boundary conditions. 
    (a) Spatially resolved magnetizations at times $tJ=0$, $10$ and $100$ for a transverse field of $g/J=0.75$. 
    The bottom three plots show (b) the imbalance, (c) the domain wall length, and (d) the entropy of entanglement across the initial interface for several transverse fields: the color coding is indicated in the color bar to the right. 
    The imbalance shows the existence of long-lived plateaus, even at transverse fields $g \approx J$. 
    All the results are shown for three different bond dimensions $\chi=181,256,362$, where the opacity increases with the bond dimension. 
    For panel (a) and (b), almost all of the data points lie on top of each other. 
}
    \label{fig1}
\end{figure}

For the purpose of studying the dynamics of quantum interfaces, we will initialize the system with two oppositely polarized magnetic domains, separated by a straight domain wall, see Fig.~\ref{fig1}a. This is a natural choice for experimental platforms, while not introducing any additional effects stemming from e.g. interface curvature.
For weak enough transverse fields, i.e., within the PXP approximation, these states remain stable due to sectors of different interface lengths being dynamically disconnected from each other, preventing thermalization up to timescales which scale exponentially with $J/g$~\footnote{Eventual thermalization is expected to happen for transverse fields $g/J\geq0.2$ as is revealed by an analysis of the levelspacing statistics of a $4\times 5$ system, see SM for more details~\cite{Supplemental}.}.

In the following we depart from any such perturbative limit and target the dynamics in strongly correlated regimes $g/J \sim 1$. 
As will be shown later, we observe pre-thermal plateaus which cannot be explained by Hilbert space fragmentation. Instead, this non-perturbative effect is related to the thermalization of the interface below a roughening temperature.

\paragraph{Numerical methods.}
In order to comprehensively explore the quantum roughening dynamics, we employ a variety of complementary tensor network techniques.
The simulations of the full dynamics on the 2D lattice were performed using TTNs and the time dependent variational principle (TDVP)~\cite{Kloss_2020, Pavesic_2024, krinitsin2025_quantumising}. 
Being loop-free, TTNs are efficiently contractable and their hierarchical structure allows for a more natural covering of the 2D lattice than the widely used matrix product states (MPS)~\cite{Silvi2019}. 
MPS will be used later in the text in order to solve time evolution of an effective 1D model. 
Furthermore, we utilize the variational uniform matrix product state algorithm (VUMPS)~\cite{Haegeman2011, Haegeman2016, Stauber2018} to study the ground-state properties of the effective model in the thermodynamic limit.
Quantum Monte Carlo (QMC) simulations~\cite{Evertz_1993,Evertz_2003} are employed to estimate effective temperatures corresponding to the domain wall initial conditions.
More details on the aforementioned methods are given in the supplementary material (SM)~\cite{Supplemental}.\nocite{Yueh2008}\nocite{Abul_Magd_2014}\nocite{Atas_2013}

\paragraph{Interface dynamics.}
A first indication for the slow thermalization dynamics is provided by the time evolution of the magnetization imbalance
\begin{equation}
I=M_T-M_B\,,
\label{eq:imbalance}
\end{equation}
with $M_{T/B}=\frac{1}{N}\sum_{i\in T/B}\sigma_i^x$ and $T/B$ indicating the top and bottom halves of the system (see Fig.~\ref{fig1}a). The imbalance is maximal in the initial state $\langle I(t=0) \rangle =1$ and it has to vanish in a thermal state.
Fig.~\ref{fig1}b shows $\braket{I(t)}$ for a range of transverse fields; the spatially resolved magnetization $\braket{\sigma_i^x(t)}$ at three time points is included exemplarily in Fig.~\ref{fig1}a.
For the largest transverse fields we expect a rapid thermalization, which is reflected by the observed rapid drop in the imbalance on a timescale of $t J \approx 1$.
The subsequent slow decay of the remaining imbalance is attributed to the diffusive approach to a finally homogeneous energy density.
The smallest field values conversely show extremely long-lived non-thermal states, due to the aforementioned emergent constraints~\cite{Balducci_2022, Balducci_2023}. 
Noticeably, however, pre-thermal plateaus seem to dominate the dynamics even up to intermediate values of the transverse fields of up to $g/J \approx 1$ and timescales of $Jt\approx10$.

The fact that the domain wall length operator $D=\frac12\sum_{\langle i,j \rangle} \left( 1 - \sigma^x_i \sigma^x_j \right)$ shown in Fig.~\ref{fig1}c strongly departs from its initial value highlights, that the existence of pre-thermal plateaus at intermediate transverse fields cannot be captured within the usual framework of Hilbert space fragmentation governed by restricted domain wall lengths.
We will instead demonstrate in the following that an effective description can be formulated in terms of a single-domain-wall approximation.

Fig.~\ref{fig1} includes a convergence check of the numerical results with different bond dimensions up to $\chi=362$. While late times and stronger transverse fields become challenging, the prominent pre-thermal plateaus are well within the regime of certain convergence. Most sensitive to varying bond dimension is the half-system entanglement entropy (Fig.~\ref{fig1}d), when splitting the system into two equal partitions along the initial interface, which shows that the moderate amounts of entanglement generated up to the lifetime of the plateaus are well captured.

\paragraph{Effective model.}

\begin{figure*}[t!]
    \centering
    \includegraphics[width=0.9\linewidth]{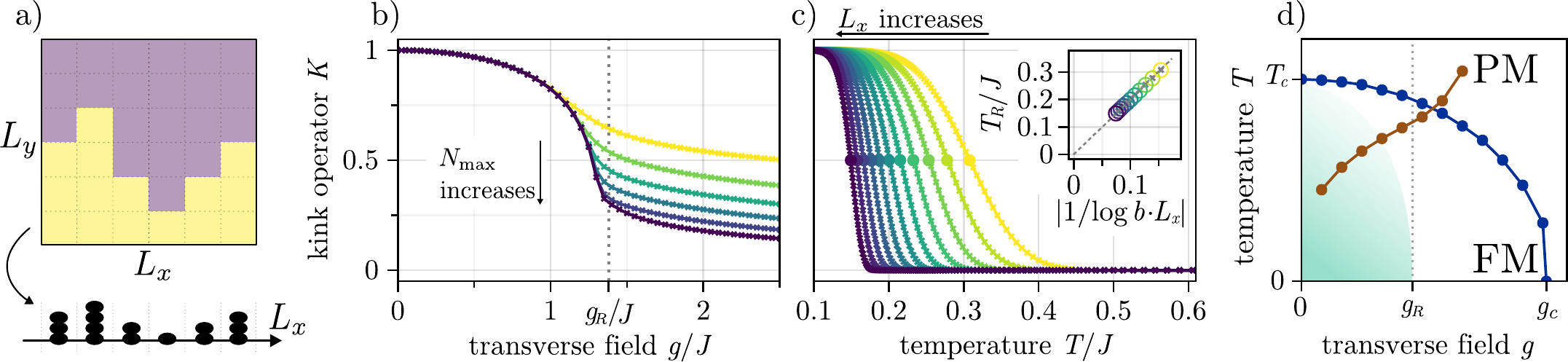}
    \caption{
    (a) Mapping from a domain wall state in the full two dimensional model to an effective one-dimensional bosonic state. 
    (b) Expectation value of the kink operator, evaluated on the groundstate of the effective model as a function of the transverse field. Different maximal occupation numbers $4 \leq N_\mathrm{max} \leq 14$ in steps of two are shown; as $N_\mathrm{max}$ increases, the kink operator starts exhibiting a jump at $g_R/J \approx 1.38$ corresponding to the roughening transition, see also the SM \cite{Supplemental}. 
    (c) Expectation value of the kink operator, evaluated on the thermal state of the effective model in the classical limit ($g=0$) as a function of temperature. We consider chain lengths $L_x \in [500,512\cdot 10^3]$, with exponentially increasing steps. The inset shows that the crossover temperature $T_R$, defined by $\langle K(T_R/J) \rangle =0.5$, vanishes in the thermodynamic limit as the inverse logarithm of the system size, confirming the analytical prediction \eqref{eq:roughening_T} which is indicated by the dashed grey line. The constant $b$ in the x-axis label is given by $b=2(1-\cos 1)/\log 2$. Data points in (b) and (c) are represented by crosses, with a linear interpolation between them. 
    (d) Phase diagram of the 2D TFIM. The blue line represents the critical line separating the ferro- from the paramagnetic phase, the red line denotes the effective temperature of the initial state based on its energy, obtained from QMC simulations, see SM~\cite{Supplemental}. Our results point towards the existence of a roughening transition at a value of $g_R/J \approx 1.38 < g_c/J$. The green shaded region shows the extended smooth interface regime, present in finite systems.
    }\label{fig2}
\end{figure*}

In order to explain the observed non-perturbative prethermal effects, we formulate an effective, one-dimensional model which corresponds to the projection of Eq.~\eqref{TFIM} onto the subspace of domain wall states with exactly one horizontal interface segment per column (see SM for a more detailed derivation~\cite{Supplemental}),
\begin{equation}
    \mathrm{H}_\mathrm{eff} = 2J \cdot \sum_{i} |N_i - N_{i+1}| - g \cdot \sum_i \left(E_i + E^\dag_i\right)\,. \label{effModel}
\end{equation}
Here, we introduced height operators $N_i$ measuring the perpendicular displacement of the domain wall, see Fig.~\ref{fig2}a. The raising (lowering) operators $E_i^\dag$ ($E_i$) are the projection of the $\sigma_i^z$, which in the manifold of single horizontal domain walls can only flip spins next to the domain wall.
The raising and lowering operators, obey the commutation relations $\left[ E_i, N_j\right] = E_i \delta_{i,j}$ similar to the standard definition used in quantum rotor models~\cite{Sachdev_2011}. Hence, the phase operator $\varphi$ defined through $E_j = \exp(i \varphi_j)$ is canonically conjugate to the height operator and the $N_i$ can alternatively be viewed as bosonic occupation number operators.
It is important to note that the model is not based on some low-order Schrieffer-Wolff transformation, but instead captures the relevant fluctuations of the initial domain wall. 

The effective model~\eqref{effModel} is also closely related to solid-on-solid-like (SOS) models~\cite{Chui_76, Beijeren_1977}, in which roughening appears as a Berezinskii–Kosterlitz–Thouless (BKT) transition~\cite{Fradkin_1978}. The critical point of $\mathrm{H}_\mathrm{eff}$ has been argued to be upper-bounded by the critical point of the quantum rotor model~\cite{Hasenfratz_1981,Sondhi1997,Hasenbusch2005}.
This is noticeably below the symmetry-breaking phase transition of the full TFIM.

Roughening is indicated by the kink operator
\begin{equation}
K_\alpha(l) = \cos(\alpha (N_1 - N_l))\,, \label{eq:kink_operator}
\end{equation}
which probes the fluctuations of the interface in the direction perpendicular to its initial orientation. It is a suited order parameter, because a value of $\braket{K_\alpha(l)}=1$ corresponds to a flat (smooth) interface, while a value of $\braket{K_\alpha(l)}=0$ corresponds to a highly fluctuating (rough) interface. 
A universal quantitative analysis would require taking the limits $\lim_{\alpha \rightarrow 0 }\lim_{l \rightarrow \infty} \langle K_\alpha(l) \rangle$ for the angle $\alpha$ and distance $l$. 
In our numerical analysis of finite systems, we choose $l=L_x$ maximal and we find that $\alpha=1$ is the minimal value, that sufficiently suppresses bulk contributions when considering the full TFIM, see SM~\cite{Supplemental} for more details. From now on we will drop the dependency on $\alpha$ and $l$, i.e., $K \equiv K_{\alpha=1}(l=L_x)$.

Fig.~\ref{fig2}b shows the ground state expectation value of the kink operator for varying $g/J$ and different values of the occupation number truncation $N_\mathrm{max}$, obtained using VUMPS.
The drop of $\braket{K}$ with increasing $g/J$ clearly indicates the transition from a smooth to a rough interface regime.
This drop becomes sharper as $N_\mathrm{max}$ is increased, pointing towards the existence of a phase transition -- a fit of the correlation length provides the critical value $g_R/J \approx 1.38$.
Further analysis strengthens the hypothesis of a BKT transition in the SOS model, see the End Matter and SM~\cite{Supplemental}.

Next, we turn towards the question wether its signatures survive even at non-zero temperatures $T$.
Note that order at $T>0$ would not violate the Mermin-Wagner theorem due to the infinite local Hilbert space dimension~\cite{Cuesta_2004}.
We consider the classical limit of the effective model, i.e., $g/J=0$, and use a transfer matrix based method to calculate the expectation value of the kink operator in the thermal state, see the End Matter and SM for more details~\cite{Supplemental}. 
The thermal expectation value of the kink operator shown in Fig.~\ref{fig2}c exhibits an extended regime with a clear signature of smooth interfaces at low temperatures for system sizes up to $L_x=5 \times 10^5$.
However, the turning point shifts with increasing system size and its location behaves perturbatively as  
\begin{equation}
    T_R/J = 2\log\left(\frac{2(1 - \cos(\alpha))L_x}{\log(2)}\right)^{-1}\,,
    \label{eq:roughening_T}
\end{equation}
see the inset of Fig.~\ref{fig2}c as well as End Matter and SM~\cite{Supplemental}. Thus, in the thermodynamic limit $L_x \rightarrow \infty$, roughening occurs immediately for any $T>0$. Nonetheless, clear signatures of a smooth interface regime at non-zero temperature survive up to very large system sizes due to the logarithmic dependence of $T_R$ on $L_x$, which has especial relevance for current experimental realizations in quantum simulators.

Fig.~\ref{fig2}d shows a sketch of the inferred phase diagram in the full two dimensional Ising model, summarizing all the previously discussed results: The ferromagnetic phase encompasses an extended smooth interface regime delimited by a roughening QPT and a system-size dependent crossover at non-vanishing temperatures. We include the effective temperatures fixed by the domain wall initial condition for the range of considered transverse fields $g/J$, indicating that signatures of a smooth interface regime will vanish already at field strengths below the critical $g_R$.
Concerning the non-equilibrium dynamics, this phase diagram suggests that the effective model can thermalize in the smooth domain wall regime, implying stability of domain walls for long times. We will show next that this prediction is even quantitatively accurate for the full dynamics of the two-dimensional TFIM.

\paragraph{Dynamical signature of roughening.} 

\begin{figure}[t!]
    \centering
    \includegraphics[width=\linewidth]{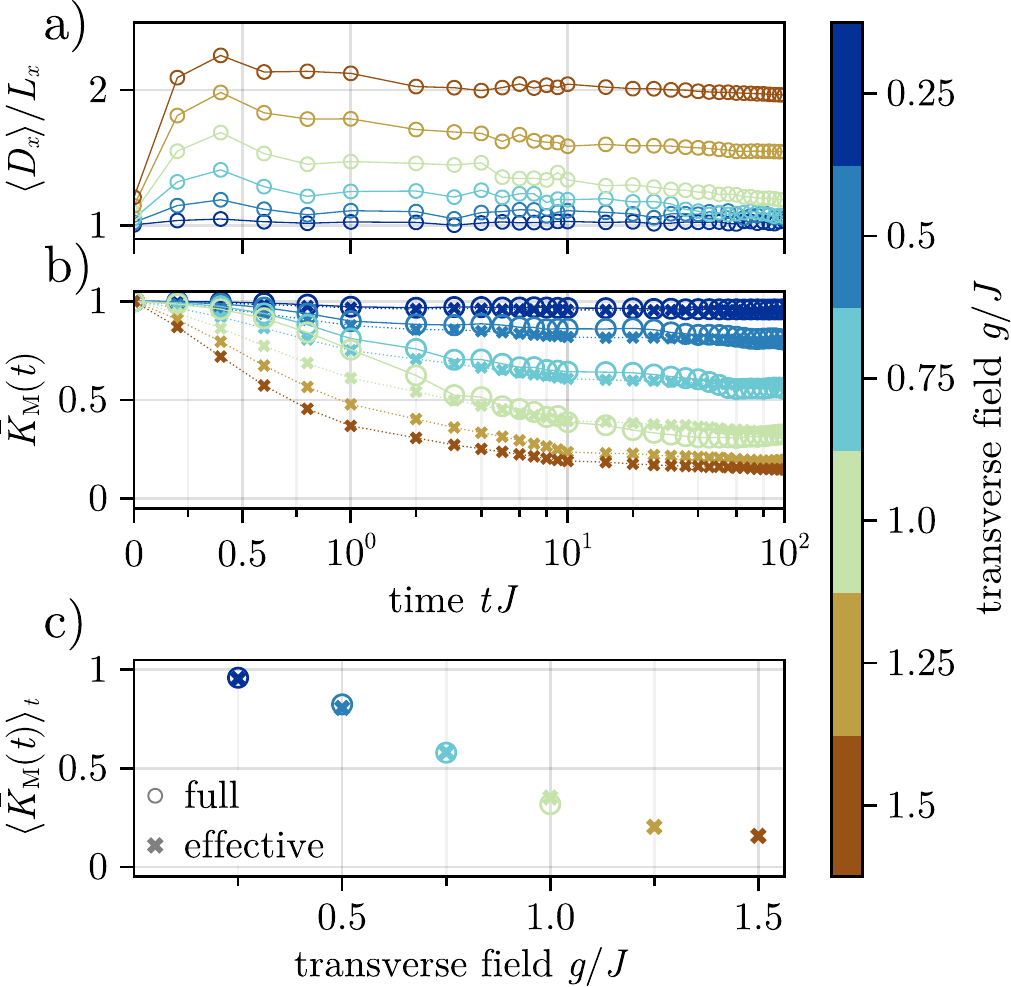}
    \caption{(a) Time evolution of the horizontal contribution to the domain wall length divided by the horizontal lattice dimension. The plot shows that up until $g/J=1$, each column has approximately one horizontal domain wall for the times considered, confirming the validity of the effective model \eqref{effModel} in that regime. 
    (b) Comparison between the effective (dashed lines/crosses) and the full model (solid lines/circles) for the time evolution of the kink operator. In order to filter out the strong fluctuations of the data we show the running mean of the kink operator. For the full model, we calculate the modified kink operator $K_\mathrm{M}= \langle K \rangle / \langle K_\mathrm{bulk} \rangle$. We don't show data for transverse fields beyond $g/J=1$, as bulk and interface contributions become increasingly difficult to disentangle. 
    (c) Late-time averages of the kink operators shown in (b), taken over the interval $tJ \in [20,100]$, as functions of the transverse field. The values agree well up to $g/J=1$, once again confirming the quantitative predictive power of the effective model in that regime.} \label{fig3} 
\end{figure}

The effective model covers the subspace of single domain-wall states without bubbles or overhang.
To check the validity of this description, we plot the time-dependence of the horizontal domain wall length $D_x=1/2 \sum_{i,j}(\unit-\sigma^x_{i,j}\sigma^x_{i,j+1})$ in Fig.~\ref{fig3}a.
The small deviations of $\langle D_x\rangle/L_x$ from one for transverse fields up to $g/J \approx 1$ support the validity of the effective model in that regime.

We now turn to a direct comparison between the dynamics of the full and the effective model -- the latter simulated using MPS.
In the full 2D TFIM, the operator probing the vertical position of the domain wall at site $i$ is given by
$N_i = \sum_{j\perp i} j/2 \cdot (\unit-\sigma^x_{i,j}\sigma^x_{i,j+1})
$
with the sum running over all lattice sites $j$ perpendicular to the domain wall. 
For a meaningful comparison with the effective model, however, we need to account for bulk contributions such as single spin flips away from the domain wall.
For that purpose, we define a modified kink operator $K_{\mathrm{M}} = \langle K \rangle / \langle K_{\mathrm{bulk}}\rangle$,
where $\langle K_{\mathrm{bulk}} \rangle$ is obtained by calculating the kink operator for a system where the interaction along the initial domain wall is removed, i.e., only bulk effects contribute to the time evolution. 
Since the time-dependent expectation values obtained from the full and the effective model differ in their high frequency fluctuations, we will moreover consider their running time-averages $\bar{K}(t)=\frac{1}{t}\int_0^tdt \langle K(t) \rangle$ instead for the direct quantitative comparison.
See the SM for more details~\cite{Supplemental}.

The time evolution of the modified kink operator in comparison with the kink operator of the effective model is shown in Fig.~\ref{fig3}b.
Data for the full model is restricted to transverse fields $g/J\leq1$, since the separation of bulk from interface effects becomes infeasible for larger fields. 
We find very good quantitative agreement between these two models, meaning that the dynamics of interfaces in the TFIM can be understood in terms of the SOS model: 
For the considered intermediate values of the transverse field, the SOS model thermalizes within the smooth interface regime (cf.~Fig.~\ref{fig2}d) and the initially flat domain wall profile remains eternally stable. 
This stability of the domain wall manifests itself in the full TFIM in the form of prethermal plateaus with $\bar K_\mathrm{M}(t)>0$. 
In contrast to the effective model, a subsequent decay of these prethermal states is to be expected in the full model, but these timescales seem out of reach for current numerical approaches.
Fig.~\ref{fig3}c displays the late-time stationary values of the running, time-averaged kink operator by taking the mean over points lying in the interval $tJ \in [20,100]$, which once more highlight the compelling quantitative agreement between the SOS model and the TFIM.
Results obtained for a 16 by 16 lattice confirm said agreement for larger system sizes, without the need to suppress temporal fluctuations via taking the running mean, see SM~\cite{Supplemental} for more details.
Finally, notice that the imbalance in Fig.~\ref{fig1}b and the kink operator in Fig.~\ref{fig3}b exhibit different lifetimes of the prethermal plateaus. 
This indicates that restoring the rotational symmetries requires longer times than the restoration of translational symmetry.

\paragraph{Discussion.}
Our analysis establishes a connection between the roughening transition and the relaxation dynamics of quantum interfaces in the 2D TFIM.
The stability of domain walls in the smooth interface regime constitutes a new mechanism for pre-thermalization beyond known ones like proximity to an integrable point~\cite{Kehrein2008} or Hilbert space fragmentation and quantum scars~\cite{Moudgalya2022}, highlighting a qualitative change within the dynamical phase diagram of the TFIM in two dimensions.

The domain wall dynamics of the quantum Ising model can be probed experimentally in state of the art Rydberg atomic systems~\cite{Zeiher_2017,Bernien2017,Scholl2021,Manovitz_2024}. 
Domain wall initial conditions can be prepared via programmable locally controlled light shifts.
and recent experiments demonstrate the feasibility of a transversely oriented magnetic field of the required intermediate magnitude~\cite{Bernien2017,Manovitz_2024}. 
Our formulation of the effective model remains unchanged for the typical anti-ferromagnetic interactions.
Since the imbalance and the kink operator are immediately accessible through snapshot measurements, we expect, that the described phenomenology is readily accessible in current Rydberg atom quantum simulators.
An exciting prospect would be the possibility to probe longer time scales in this way.

Furthermore, it will be interesting to investigate other initial domain wall configurations in the light of roughening.
Examples include the observed self-straightening dynamics of a initial zig-zag configuration~\cite{Manovitz_2024} and potential implications for false vacuum decay probed via bubble formation~\cite{Coleman_1977,Lagnese2021,Milsted2022,Zenesini2024}.
More generally, the impact of curvature on the phenomenology of the roughening dynamics remains to be explored in a future work.
Another immediate question concerns the generalization to other symmetries, since the results presented here rely on a $\mathbb Z_2$ symmetry-broken phase. 

\begin{acknowledgments}
\paragraph{Acknowledgements.}
We acknowledge fruitful discussions with A. Segner, J.-D. Urbina, M. Steinhuber, L. Behringer, S. Tomsovic, T. Mendes-Santos, A. Gambassi, U.-J. Wiese.
The tree tensor network simulations presented in this work were produced with \rm{TTN.jl}~\cite{Tausendpfund2024}, a software package we developed based on the \rm{ITensor} library~\cite{Fishmann2022}. The MPS simulations were performed with the help of \rm{ITensor}. The QMC simulations were produced with the help of the \rm{ALPS} software package~\cite{Bauer_2011}.
MS and WK were supported through the Helmholtz Initiative and Networking Fund, Grant No. VH-NG-1711.
We acknowledge support from the Deutsche Forschungsgemeinschaft (DFG) under 
Germany's Excellence Strategy - Cluster of Excellence Matter and Light for Quantum Computing (ML4Q) EXC 2004\slash 1 – 390534769 (NT, MR and MS),
under Grant No. 277101999 -- CRC network TR 183 (NT and MR),
under project 499180199 -- FOR 5522 (MH).
MH has received funding from the European Research Council (ERC) under the European Union’s Horizon 2020 research and innovation programme (grant agreement No. 853443).
MR and NT acknowledge funding under Horizon Europe programme HORIZON-CL4-2022-QUANTUM-02-SGA via the project 101113690 (PASQuanS2.1).
The authors gratefully acknowledge the Gauss Centre for Supercomputing e.V. (www.gauss-centre.eu) for funding this project by providing computing time through the John von Neumann Institute for Computing (NIC) on the GCS Supercomputer JUWELS~\cite{JUWELS} and through FZJ on JURECA~\cite{JURECA2021} at J\"ulich Supercomputing Centre (JSC).
Data and code are available at~\cite{krinitsin_2025}.

\end{acknowledgments}


\bibliography{refs}

\appendix

\section{End Matter}
\paragraph{BKT transition of the SOS Model.}
In this paragraph, we demonstrate that the nature of the transition in the effective SOS model~\eqref{effModel} belongs to the BKT universality class.
To this end, we use the VUMPS algorithm~\cite{Haegeman2011,Haegeman2016,Stauber2018} to study the groundstate properties of the effective model. This allows us to work directly in the thermodynamic limit with a fixed truncation of bosonic excitations $\nmax \in 2\Z$.
In the two-dimensional model, this is equivalent to considering an infinitely long slab of width $L_y = \nmax + 1$ with the interface oriented along the infinite direction.
To observe the critical properties, it is necessary to scale the results against $\nmax \to \infty$.

In practice, we considered $\nmax\le 14$.
Since the local Hilbert space scales as $\dim(\mathcal{H}_{\rm loc}) = \nmax +1$, a maximum bosonic occupation of fourteen already leads to very long simulation times of $13\times10^3$ seconds per iteration for a bond dimension of $\chi = 600$.

\begin{figure}[t!]
    \centering
    \includegraphics[width=\linewidth]{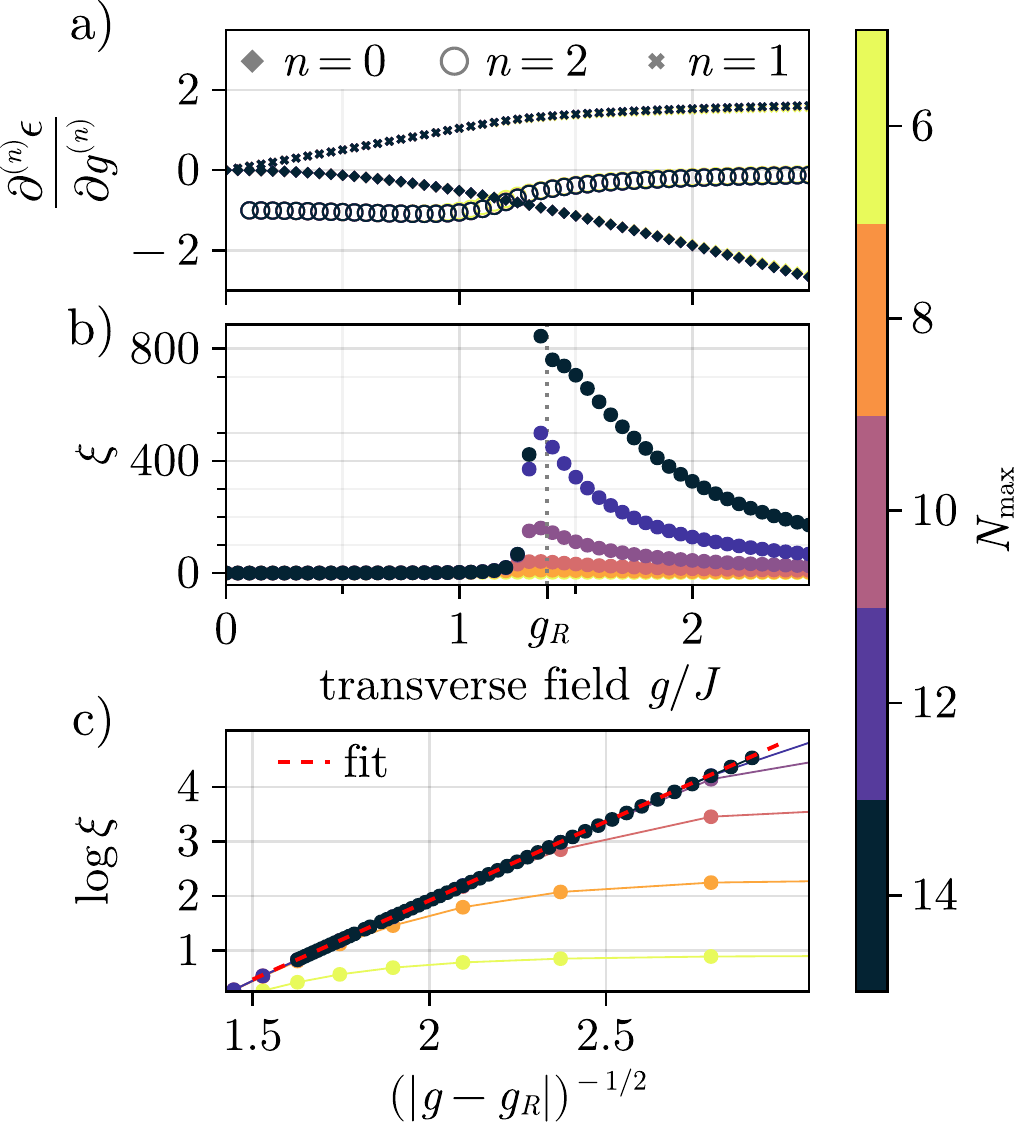}
    \caption{a) The energy density and its first two derivatives with respect to the transverse field $g$, as a function of g. All of the different order do not show any sign of non-analyticity around the transition point, excluding the possibility of it being a first-/second order phase transition. b) The correlation length shows a divergence at $g=g_R$ when increasing the maximal bosonic occupation number. All results are obtained with a bond dimension of $\chi = 600$ and are converged for $\nmax < 14$. For the $\nmax = 14$, the points around the transition are not yet fully converged showing numerical artefacts for $g>g_R$. c) Fit of \eqref{eq:corr_length} to the data, showing the correct BKT type divergence law of the correlation length. Increasing $\nmax$ again leads to a better agreement between the data and the expected law. The critical transverse field obtained from the fit for the largest $\nmax=14$ is $g_R/J=1.38$. } \label{endMatter} 
\end{figure}

In contrast to continuous phase transitions described by the Landau-Ginzburg theory, the BKT transition cannot be detected by the divergence of any derivative of the energy density $\epsilon(g)$, which is an analytical function of $g$.
This is demonstrated in Fig.~\ref{endMatter}a via the example of the first and second derivative of $\epsilon(g)$.
Moreover, we observe that $\epsilon(g)$ and its derivatives are quickly converging in $\nmax$.

Characteristic of the BKT transition is the exponential divergence of the correlation length when approaching the critical value $g_R$,
\begin{equation}
    \xi(g) = \xi_0 \exp\left(\frac{B}{\sqrt{|g - g_R|}}\right), \ \mathrm{for} \ g<g_R\,. \label{eq:corr_length}
\end{equation}
Here, $\xi_0$ and $B$ are non-universal constants depending on the model and on the observable used to extract the correlation length.
We observed the largest correlation length for the vortex-vortex correlation function
\begin{equation*}
    C(l) = \braket{e^{-iN_0} e^{iN_l}}\xrightarrow[]{l\to\infty} Ae^{-\frac{l}{\xi}} + C_\infty
\end{equation*}
which is closely related to the kink operator~\eqref{eq:kink_operator}. The constant $C_\infty$ is expected to be close to one in the $g<g_R$ region, while we expect it to vanish after the BKT transition for $\nmax \to \infty$. In particular one has $\braket{K_\alpha(l)} \to |C_\infty|$ for $l \to \infty$.
The results are displayed in Fig.~\ref{endMatter}b for $6\leq\nmax\leq14$. For increasing $\nmax$, the correlation length shows a divergent behavior around $g/J \approx 1.4$ with $\xi \sim 800$ for $\nmax = 14$.

To quantify that $\xi(g)$ has the correct behavior approaching $g_R/J$, we fit $\xi(g)$ for $\nmax = 14$ for $1.1<g/J<1.385$ with higher resolution using Eq.~\eqref{eq:corr_length}. From this fit, we obtain $g_R/J \approx 1.38$.
In Fig.~\ref{endMatter}c we plot the logarithmic correlation length as a function of $1/\sqrt{|g- g_R|}$ for $g<g_R$ and $6\leq\nmax\leq14$ using the value of $g_R$ obtained by the fit.
Increasing $\nmax$ leads to a better agreement between the data and the prediction for the correlation length, which should be exact in the limit $\nmax \to \infty$. For lower $\nmax$ the curve starts flattening out and thus deviating sooner, illustrating the finite $\nmax$ effects.

\paragraph{Finite temperature crossover in the classical limit.}

In the classical limit $g\to 0$ the effective model~\eqref{effModel} contains only commuting operators $|N_j - N_{j+1}|$. Thus, all eigenstates are simply given by product states $\ket{\lbrace s_j \rbrace}$, fixing the occupation on the $j$-th site.
To study the thermodynamic properties in this limit, it is sufficient to replace the number operator $N_j$ by its eigenvalues $s_j \in \lbrace 0,\nmax\rbrace$
\begin{equation}
    \mathcal{H}_\mathrm{class}(\lbrace s_j \rbrace) = 2J \sum_{j=1}^{L_x-1} |s_j-s_{j+1}| = \sum_{j=1}^{L_x-1} h(s_j, s_{j+1})\,.\label{eq:eff_class_model}
\end{equation}
Let $V_{s, s^\prime} = \exp(-\beta h(s,s^\prime))$ be the transfermatrix of the classical system and define the general $\alpha$ twisted boundary vector $\ket{E(\alpha)} = \sum_{s=0}^\nmax e^{i\alpha s} \ket{s}$. 
The $\alpha$ twisted partition function of a chain of length $L_x$ with open boundary conditions can be written compactly as:
\begin{equation}
    \begin{split}
    \mathcal{Z}(\alpha) &= \sum_{s_1=0}^\nmax ...\sum_{s_{L_x}=0}^\nmax e^{-\beta \mathcal{H}_{\rm class}(\lbrace s_j \rbrace) - i\alpha(s_1 - s_{L_x})} \\
    {}& = \braket{E(\alpha)|V^{L_x-1}|E(\alpha)}\, .
    \end{split}
\end{equation}

The standard partition function corresponds to zero twisting, i.e., $\mathcal{Z} = \mathcal{Z}(0)$. We can now write down the expectation value of the string operator $\braket{K_l}$.
For simplicity, we consider $l = L_x$, i.e., the end-to-end expectation value. 
Using the twisted partition function, we obtain
\[
\braket{K_\alpha(L_x)} = \frac{\mathcal{Z}(\alpha)}{\mathcal{Z}(0)}
\]

The calculation can be further simplified by diagonalizing the transfer matrix $V = U \Lambda U^\dag$. In this case, the twisted partition function can be written as
\[
\mathcal{Z}(\alpha) = \lambda_1^{L_x-1} \sum_{s = 0}^\nmax f_n^{L_x-1} |c_n(\alpha)|^2
\]
where $\lambda_1$ is the largest eigenvalue, $f_n = \lambda_n/\lambda_1$, 
and $c_n(\alpha)$ are the form factors obtained by calculating the overlap between the eigenstates $\ket{\psi_n}$ of $V$ and the twisted boundary vector $\ket{E(\alpha}$.

Numerically we found the operator $\braket{K_\alpha(L_x)}$ to rapidly converge in $\nmax$, more specifically $\nmax = 200$ is large enough for our analysis, see SM~\cite{Supplemental} for more details.

It is also possible to understand the crossover behavior analytically and derive a functional form for the transition temperature $T_R(L_x)$.
Using the exact eigenvectors and eigenvalues of the transfer matrix $V$ of the classical model~\eqref{eq:eff_class_model}, we were able to find an approximation of the end to end string operator in the limit of large $\nmax,L_x\gg1$, and small $q\ll 1$~\cite{Supplemental}:
\begin{align}
    \braket{K_\alpha(L_x)} \approx \exp\left(-2q(1-\cos(\alpha)L_x\right)\, .
\end{align}
By setting $\braket{K_\alpha(L_x)} = 1/2$ and solving for $T/J = -2/\log(q)$, we find that the transition temperature $T_R$ should vanish logarithmically with the system size $L_x$:
\begin{equation}
\label{eq:transition_temp}
T_R(L_x)/J = 2\log\left(\frac{2(1 - \cos(\alpha))L_x}{\log(2)}\right)^{-1}\,,
\end{equation}
which has been confirmed numerically, see inset of Fig.~\ref{fig2}c.


\setcounter{equation}{0}
\onecolumngrid
\newpage
\section*{Supplemental materials for ``Roughening dynamics of interfaces in the two-dimensional quantum Ising model''}
\input{SM.tex}

\end{document}

%% file: SM.tex
\renewcommand{\theequation}{S\arabic{equation}}
\renewcommand{\thefigure}{S\arabic{figure}}
\newcommand{\revision}[1]{\textcolor{blue}{#1}}

\section{Tree Tensor Networks and Time Dependent Variational Principle}
\begin{figure}[h!]
    \centering
    \includegraphics[width=0.5\linewidth]{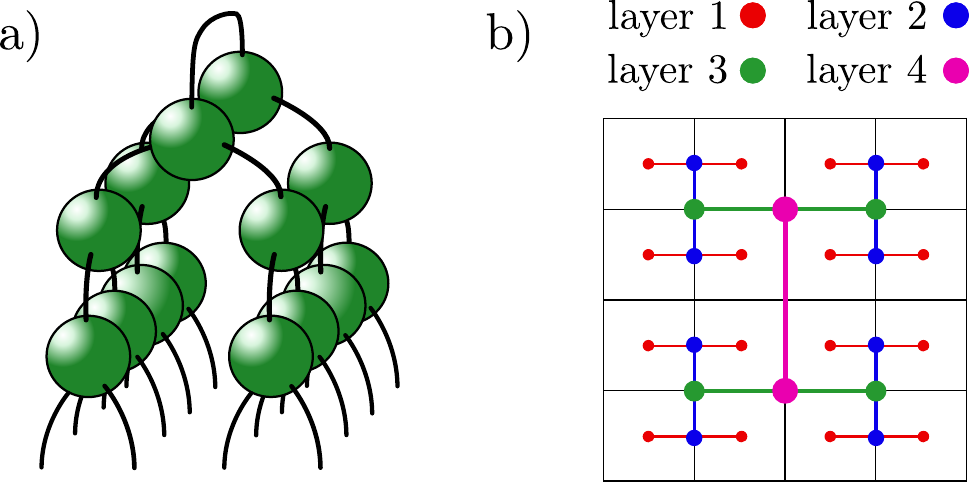}
    \caption{a) Visualization of a binary tree tensor network and its connectivity on a 2D lattice. Going horizontally from layer to layer changes the direction of the connection. This is presented more clearly in b) which shows said direction as a top-down view on the lattice, with different colors representing different layers.} \label{fig:fig_ttn} 
\end{figure}
We use binary tree tensor networks, see Fig.~\ref{fig:fig_ttn}~a for a graphical representation thereof, in order to simulate the dynamics of a quantum system on a two dimensional lattice. By alternating the x-/y-direction for the link connection in each layer of the tree one can achieve a more natural cover of the lattice, see Fig.~\ref{fig:fig_ttn}~b. 
The orientation of the tree tensor network with respect to the lattice is such that the top node connects the two halves of the systems across the interface. The choice of the orientation might play an important role for rotationally asymmetric initial states such as ours. We checked the effect of that choice on the results by rotating the tensor network structure by $90$ degrees. This did not have any effect for the system sizes and bond dimensions, for the intermediate transverse fields ($g/J \leq 1$) considered.

An advantage of such a hierarchical structure lies in the fact that two spins at distance $l$ on the two dimensional lattice are connected by at most $O_\mathrm{ttn} \sim \log l$ tensors, in contrast to the $O_\mathrm{mps} \sim l$ tensors needed in an MPS. The absence of loops means that one can define an isometry center and thus has access to efficient algorithms such as the TDVP algorithm~\cite{Kloss_2020} used here.
Its 1-site version ensures that the energy and norm of the state are conserved. However as a downside, it doesn’t allow for the bond dimension to be adapted throughout the algorithm. Thus, the desired bond dimension needs to be encoded in the initial state from the beginning, which is done by padding the product state with zeros. 
For the time step we chose $dt=0.1$.

The TTN structure allows one to easily access the von Neumann entanglement entropy, by performing an SVD on the tensor which splits the system into the desired partitions — the entanglement entropy is given in terms of the singular values $\lambda_i$: $S = - \sum_i \lambda_i ^2 \log \lambda_i^2$.

Convergence is checked by comparing results obtained at different bond dimensions. In our case, local observables often show rapid convergence in terms of bond dimensions, while the entanglement entropy carries information about all higher order correlations, and is a more sensitive measure for convergence, see Fig.~1d in the main text.

We work with open boundary conditions (OBC) in both lattice directions due to the slower buildup of entanglement compared to periodic boundary conditions (PBC). In principle, periodic boundary conditions would be preferable in the direction parallel to the interface to reduce boundary effects. 


Another consequence of the binary tree structure is that the tensors appearing in the tree are of size $\chi^2 \times \chi$ and thus of the skinny-tall type, which allows for an efficient parallelization of the QR decompositions on these tensors. Since tensor contractions are already naturally parallelizable, this means that the overall runtime of the simulations profit substantially from GPU acceleration. In our case, we observe a speedup of around $70\times$ over CPU performance. More specifically, the most expensive simulation at a bond dimension $\chi=362$, at a time step $dt=0.1$ up to $t_\mathrm{max}=100$ takes around two days on a single \emph{A100} GPU -- on a CPU (\emph{Xeon Gold}) the same calculation would take around 145 days.

\section{Effective model}

In the following section we present and motivate a derivation of the effective model. The domain wall length operator, introduced in the main text, can be split into a horizontal and vertical component, by performing the sum only over horizontally
\begin{equation}
D_x=\frac{1}{2} \sum_{i,j}\left(\unit-\sigma^x_{i,j}\sigma^x_{i,j+1}\right)
\end{equation}
or vertically
\begin{equation}
D_y=\frac{1}{2} \sum_{i,j}\left(\unit-\sigma^x_{i,j}\sigma^x_{i+1,j}\right)
\end{equation}
oriented bonds, respectively.
With that, the transverse field Ising Hamiltonian can be rewritten as
\begin{equation}
    \mathrm{H} = 2 J D_x + 2 J D_y - g \cdot \sum_i \sigma^z_i\,,
\end{equation}
which coincides with the original Hamiltonian up to an irrelevant, additive constant.
The effective model is introduced as the projection onto the subspace of domain wall states that have exactly one horizontal domain wall per column, thus fixing $D_x/L_x=1$. 
As a consequence these states cannot exhibit overhangs or bubbles of oppositely oriented spins within the bulk.

The vertical contribution to the domain wall length operator is determined by the difference of the vertical position $N_i, \, i\in \{1,...,L_x\}$ of two neighboring segments
\begin{equation}
    D_y \propto \sum_i|N_i-N_{i-1}|\,,
\end{equation}
while the absolute value reflects that the result is independent of the relative ordering in each pair.

\section{Kink operator}
\begin{figure}[h!]
    \centering
    \includegraphics[width=\linewidth]{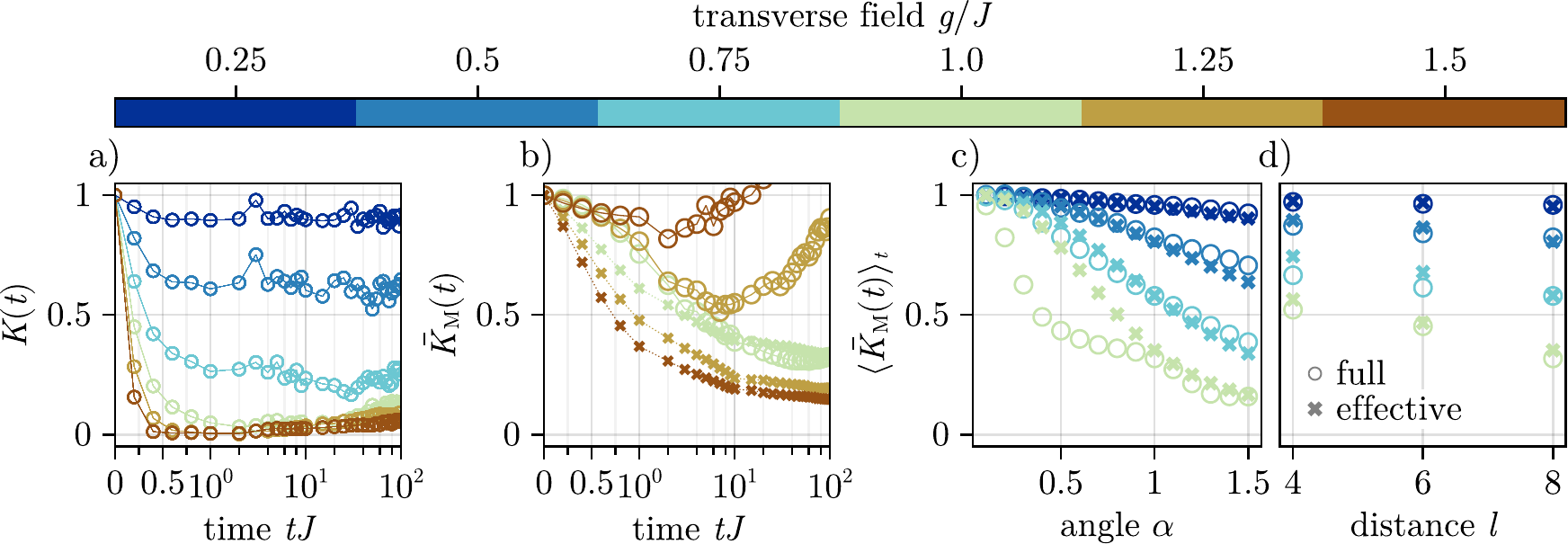}
    \caption{a) Time evolution of the bare, i.e. unmodified kink operator in the 2D TFIM. b) Time evolution of the modified kink operator for transverse fields $g/J=1,1.25,1.5$, for both the 2D TFIM and 1D effective model. The plot shows unphysical growth of the modified kink operator for $g/J > 1$, signaling the breakdown of the assumption that bulk and interface effects can be separated. Plots c) and d) show the modified, kink operator, averaged over the interval $tJ \in [20,100]$, as a function of c) the angle $\alpha$ and d) the distance $l$. The distance does not have much influence on how well the effective and full model agree, which is why throughout the work we chose the largest distance possible for our lattice, i.e. $l=8$. This is different for the angle where we observe larger deviations for the largest transverse field $g/J=1$. The smallest angle which showed the smallest deviations across all values of the transverse fields considered is $\alpha=1$, which we used throughout this work.} \label{fig:fig_kink} 
\end{figure}

The kink operator measures by how much the interface fluctuates across its length. In the case of a smooth interface, the difference in the heights of the interface at different points of the domain wall will be close to 0, hence $K=1$, while for a rough interface, the cosine function ensures that large fluctuations add up to $K=0$. This behavior is reflected in Fig.~\ref{fig:fig_kink}~a, which shows how the kink operator stays close to $K=1$ during time evolution up to $tJ=100$ for small transverse fields, while for large ones it drops to zero on a timescale of $0.1 \leq tJ \leq 1$. This as well as all subsequent simulations in that section were performed on an $8\times8$ lattice.

Since the expectation value of the kink operator is real, we can alternatively write it as the real part of the complex exponential instead of the cosine, i.e. 
\begin{equation}
    K_\alpha(l) = \exp(-i \, \alpha (N_1 - N_l)). \label{eq:kink_operator_exp}
\end{equation}
From that, the modified kink operator $K_{\mathrm{M}} = K / K_{\mathrm{bulk}}$,  introduced in the main text, can be understood as splitting the bulk and interface contributions in the exponent into a sum, which in turn allows one to factor out the bulk contribution in the calculation of the kink operator. This separation only works when neglecting cross terms introduced by the exponential function, i.e. when the bulk and interface contributions can be clearly separated. This becomes infeasible for transverse fields $g/J>1$ as can be seen in Fig.~\ref{fig:fig_kink}b, where the kink operator first drops and then goes up again in time for $g/J\geq1.25$. We want to stress that the observed rising of the kink operator an artifact of the method and not a physical effect. 

The angle $\alpha$ and distance $l$, appearing in the calculation of the kink operator allow to tune its sensitivity. In Fig.~\ref{fig:fig_kink}c and d we show their influence on the time average of the running mean of the kink operator. 

As expected, $K \rightarrow 1$ for $\alpha \rightarrow 0$. For transverse fields $g/J<1$ we observe little deviation between the results for the effective and full model. For $g/J=1$ this is not the case anymore, which is to be expected as we are approaching the roughening transition. For the system sizes considered in this work, we found the value of $\alpha=1.0$ to provide the best agreement between the effective and full model across all transverse fields of interest, and thus the highest sensitivity to the roughening transition. This can be understood as smaller angles leading to an overall higher sensitivity to boundary and unwanted bulk effects, which due to their large vertical distance from the initial interface provide a big contribution to the kink operator, which in turn is detrimental for the system sizes considered. A larger value of $\alpha$ enhances all contributions to the kink operator, while boundary effects are being "washed" out due to the cosine function, which will add up larger arguments to 0 due to the rapid sign changes. One thus needs to be careful to not make the angle too large, otherwise the contributions coming from the interface will be lost as well.

Figure~\ref{fig:fig_kink}~d shows the effect of changing the distance to $l$ -- the effective and full model agree for all transverse fields and distances $l=4,6,8$ shown here.

Results for the kink operator on a 16 by 16 lattice are provided in Figure~\ref{fig:fig_kinkL16}, in a similar fashion to what has been presented in the main text for an 8 by 8 lattice. 
Due to the larger overall entanglement, the simulations only provide reliable results up to a timescale of $t_\mathrm{max} = 30$. 

Figure~\ref{fig:fig_kinkL16}~a) shows the domain wall length contribution coming from horizontally aligned interface lengths, normalized by the length of the lattice in x-direction. 
Once again this confirms that up to $g/J=1$, the number of domain walls per column does not scale extensively with the vertical lattice size.
The agreement between the time evolution of the modified kink operator calculated in the full 2D and effective models is shown in Figure~\ref{fig:fig_kinkL16}~b).

We want to emphasize that for that comparison we did not utilize the running mean of the modified kink operator as has been done in the main text for the 8 by 8 system. 
This is because compared to the 8 by 8 system, the kink operator calculated in the 16 by 16 system does not show strong temporal fluctuations, allowing for a direct comparison.

The agreement between the two models is made even more apparent when comparing the late time plateau values of the respective kink operators, obtained by taking the mean points within the region $tJ \in [5,30]$, see Figure~\ref{fig:fig_kinkL16}~c).

\begin{figure}[h!]
    \centering
    \includegraphics[width=0.6\linewidth]{figures/supplemental_material/sup_figure_kink_operatorL16.pdf}
    \caption{Kink operator for a 16 by 16 system, evolved up to a time of $t_{\max}=30$. a) Time evolution of the horizontal contribution to the domain wall length divided by the horizontal lattice dimension. The plot shows that the number of horizontal domain walls in each column does not scale extensively with the vertical system size, confirming the validity of the effective model in that regime. (b) Comparison between the effective (dashed lines/crosses) and the full model (solid lines/circles) for the time evolution of the kink operator. For the full model, we calculate the modified kink operator $K_\mathrm{M}= \langle K \rangle / \langle K_\mathrm{bulk} \rangle$. We don't show data for transverse fields beyond $g/J=1$, as bulk and interface contributions become increasingly difficult to disentangle. (c) Late-time averages over the interval $tJ \in [5,30]$ of the kink operators shown in (b) as functions of the transverse field.  The values agree well up to $g/J=1$, once again confirming the quantitative predictive power of the effective in that regime.} \label{fig:fig_kinkL16}
\end{figure}

\section{Exact Solution of the Classical Limit and Approximations}

The thermodynamic properties at temperature $T = 1/\beta$ of the SOS model in the classical limit $g\to 0$ are defined by the transfer-matrix
\begin{equation}\label{eq:transfermatrix_supp}
    V_{s, s^\prime} = \exp(-\beta h(s,s^\prime)) = q^{|s - s^\prime|} \, ,
\end{equation}
with $q = \exp(-2J\beta)$ and $s\in \lbrace 1,\dots, M\rbrace$. Here, we have shifted the range of values for $s$ by one with respect to the main text and defined $M = \nmax + 1$ to simplify the notation in the following discussion.
We are interested in evaluating the end-to-end kink correlation function given by
\[
\braket{K_\alpha(L_x)} = \frac{\mathcal{Z}(\alpha)}{\mathcal{Z}(0)} 
\]
with the twisted partition function
\begin{equation}\label{eq:twst_prtn}
\begin{split}
\mathcal{Z}(\alpha) &= \lambda_1^{L_x-1} \sum_{n = 1}^M f_n^{L_x-1} |c_n(\alpha)|^2\,,\\
c_n(\alpha) &= \sum_{s = 1}^M e^{i\alpha s}\psi_n(s)\,,\quad f_n = \frac{\lambda_n}{\lambda_1} \,,
\end{split}
\end{equation}
where $\ket{\psi_n} = \sum_{s=1}^M \psi_n(s)\ket{s}$ are the eigenvectors of $V$ and $\lambda_n$ the corresponding eigenvalues with $\lambda_1$ being the largest eigenvalue.

The inverse of Eq.~\eqref{eq:transfermatrix_supp} is connected to a single-particle tight-binding Hamiltonian $H_{\rm sp}$ via
\begin{equation}\label{eq:sp_ham}
H_{\rm sp} = (1 - q^2) V^{-1} - (1 + q^2) \unit =-q\,\sum_{s=1}^{M} \left(\ket{s}\bra{s+1} + \hc\right) - q^2\left(\ket{0}\bra{0} + \ket{M}\bra{M}\right)\, ,
\end{equation}
which is a standard next nearest neighbor hopping Hamiltonian with additional edge potentials.
The eigenvalue and eigenvectors of this tight binding model are known~\cite{Yueh2008}
\begin{equation}\label{eq:eigen_sp}
    \psi_n(s) = \mathcal{N}\left\lbrace\sin(s\theta_n) - q \sin((s-1)\theta_n)\right\rbrace\,,\ 
    \epsilon_n = -2q\cos(\theta_n) \, ,
\end{equation}
where $\ket{\psi_n} = \sum_s \psi_n(s) \ket{s}$ are the eigenvectors, $\mathcal{N}$ a normalization factor, and the $\theta_n$ are defined by the roots of the equation
\begin{equation}\label{eq:theta_root_eq}
    \sin((M+1)\theta_n) + q^2\sin((M-1)\theta_n) - 2q\sin(M\theta_n) = 0 \,,
\end{equation}
with $\theta_n \in [0,\pi]$.
For $M\to \infty$ and $q\to0$, the solutions of this equation become dense in the interval $[0,\pi]$, with the distance between two solutions vanishing as $|\theta_{n+1} - \theta_n| \sim 1/M$. In first order one finds $\theta_n = n\pi/(M+1)$ similar to a homogeneous chain without the additional edge potentials.

By inverting Eq.~\eqref{eq:sp_ham}, the eigenvalues of $V$ are given by \[
\lambda_n = \frac{1 - q^2}{1 + q^2 - 2q \cos(\theta_n)} \, ,
\]
while the eigenvectors are still given by $\psi_n(s)$ from Eq.~\eqref{eq:eigen_sp}. 

From the exact expression of the eigenvalues and eigenvectors, we can now proceed to calculate the form factors $c_n(\alpha)$.
Note that due to the reflection symmetry of the Hamiltonian~\eqref{eq:eigen_sp} one has $\psi_n(M-s) = (-1)^n\psi_n(s)$ and thus $|c_n(\alpha)|^2 = |c_n(2\pi- \alpha)|^2$.
Thus it is sufficient to restrict the analysis to $\alpha \in [0,\pi]$.

Using the summation identity
\[
    \sum_{s=1}^M e^{i\alpha s} \sin(s\theta + \delta) = \frac{e^{i\alpha}}{2i}
    \left(
        e^{i\frac{\theta(M+1) + 2\delta}{2} } g_M(\alpha + \theta)
    -  e^{-i\frac{\theta(M+1) + 2\delta}{2} } g_M(\alpha - \theta)
    \right)\,,\ 
    g_M(x) = \frac{\sin\left(
            \frac{x}{2}M
        \right)}{\sin\left(
            \frac{x}{2}
        \right)}
\]
derived from the finite geometric sum, we find an exact expression for the form factors

\begin{equation}
\begin{split}
    |c_n(\alpha)|^2/\mathcal{N}^2 = &\frac{g_M(\theta_n + \alpha)^2 + g_M(\theta_n - \alpha)^2}{4}
    \left(
        1 + q^2 - 2q \cos(\theta_n)
    \right)\\
    {}& - \frac{g_M(\theta_n + \alpha)g_M(\theta_n - \alpha)}{2}\left(
        \cos(\theta_n(M+1)) + q^2 \cos(\theta_n(M-1)) - 2q\cos(\theta_n M)
    \right) \, .
\end{split}
\end{equation}

Note that $c_n(\alpha)$ has a unique singularity at $\theta_\star = \alpha$ coming from the function $g_M(\theta - \alpha)$ for $\theta$, $\alpha \in [0,\pi]$.
Let $n_\star$ be the solution to Eq.~\eqref{eq:theta_root_eq} with $\theta_{n_\star}$ close to $\theta_\star$.

The most dominant contribution to the twisted partition function $\mathcal{Z}(\alpha)$ is thus given by the $n_\star$ form factor in the large $M$ limit
\[
\mathcal{Z}(\alpha)/\lambda_1^{L_x-1}\approx f_{n_\star}^{L_x-1} |c_{n_\star}(\alpha)|^2 \, .
\]

Since the distance between two solutions vanishes as $1/M$ in the large $M$ limit, we expect $|\theta_{n_\star} - \theta_\star| \approx \delta/M$, where $\delta$ is a constant of order one and independent of $\alpha$.
Expanding $c_{n_\star}$ for small $\delta/M$ and keeping only the dominant part, we arrive at
\[
|c_{n_\star}|^2/\mathcal{N}^2 \approx \frac{M^2}{4}\frac{\sin(\delta/2)^2}{(\delta/2)^2}(1 + q^2 - 2q\cos(\alpha))\,,
\]
which diverges as expected for $M\to\infty$.

Next consider the weights $f_n$ and expand them around $n_\star$
\[
f_{n_\star} = \frac{\lambda_n}{\lambda_1} = \frac{1 + q^2 - 2q \cos(\theta_1)}{1 + q^2-2q\cos(\theta_{n_\star})} \approx 
\frac{\Gamma}{(1 - \frac{2q}{1+q^2}\cos(\alpha))}\,.
\]

Here, $ \Gamma = \frac{1 + q^2 -2q\cos(\theta_1)}{1 + q^2}$ is a $\alpha$ independent constant.

Assuming $q$ small one finds in the $L_x\to \infty$ limit
\[
 f_{n_\star}^{L_x -1} \approx \Gamma \left(1 + \frac{2q}{1+q^2}\cos(\alpha)\right)^{L_x} \approx \Gamma e^{\frac{2q}{1+q^2}\cos(\alpha)L_x} \, .
\]

Combining the results, we obtain for the end-to-end kink correlator
\begin{equation}\label{eq:supp_class_kink_op}
    \braket{K_\alpha(L_x)} = \frac{\mathcal{Z}(\alpha)}{\mathcal{Z}(0)}
    \approx \frac{1 + q^2 - 2q\cos(\alpha)}{(1 - q)^2}
    \exp\left(-\frac{2q}{1+q^2}(1 - \cos(\alpha))L_x\right) \approx e^{-2q(1-\cos(\alpha))L_x} \, ,
\end{equation}
where the last approximation is valid in the small $q$ and large $L_x$ limit.
Note that $1 -\cos(\alpha) > 0$ for $\alpha \in (0,\pi)$, such that $\lim_{L_x\to\infty}\braket{K_\alpha(L_x)} =0$ for all values of $\alpha \in (0,\pi)$.

\begin{figure}[h!]
    \centering
    \includegraphics[width=0.9\linewidth]{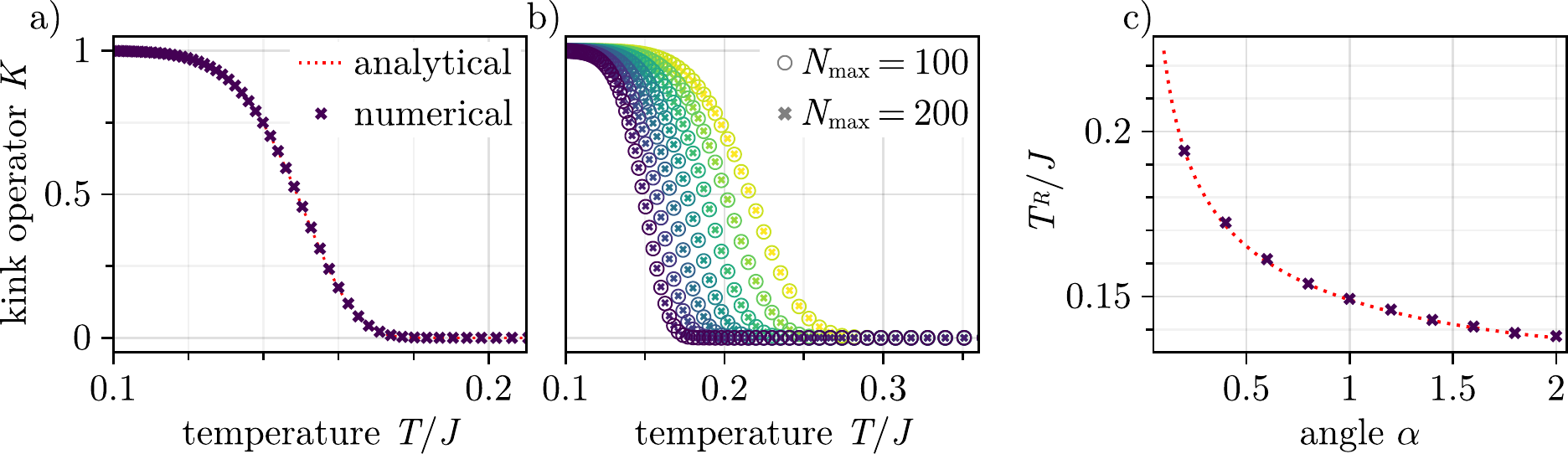}
    \caption{a) Kink operator as a function of the temperature for system size $L_x=512 \cdot 10^3$ and $\nmax = 200$. The numerically obtained data points agree very well with the analytical prediction~\eqref{eq:supp_class_kink_op}. b) Results for the kink operator as a function of $T$ for different system sizes and maximal bosonic occupation numbers. System sizes considered here lie in $L_x \in [500,512\cdot 10^3]$, with exponentially increasing steps. The data points for $N_{\rm max}=100,200$ lie on top of each other, meaning results are converged in the maximal bosonic occupation number. c) Dependence of the transition temperature on the angle $\alpha$ at $L_x=512 \cdot 10^3$, following the dependence~\eqref{eq:supp_transition_temp} (grey, dotted line), obtained from analytical considerations on the transfermatrix. 
    } \label{fig:fig_trm} 
\end{figure}

Fig.~\ref{fig:fig_trm}~a shows the agreement between numerical data and the analytical expectation~\eqref{eq:supp_class_kink_op} for $L_x=512 \cdot 10^3$ and $\nmax = 200$, confirming the approach presented here. In Fig.~\ref{fig:fig_trm}~b we numerically test the dependence of the results on the maximal bosonic occupation number $\nmax$ for different, color-coded system sizes, establishing that $\nmax = 200$ is sufficient to reliably calculate the kink operator for the system sizes considered here.

We define the transition temperature in the classical limit by setting $\braket{K_\alpha(L_x)} = 1/2$, and solving for $T/J = -2/\log(q)$:
\begin{equation}\label{eq:supp_transition_temp}
T_R(L_x) = 2\log\left(\frac{2(1 - \cos(\alpha))L_x}{\log(2)}\right)^{-1}\,.
\end{equation}
This relation is tested numerically by calculating the end-to-end kink correlator both as a function of the system size for a fixed $\alpha$, see inset of Fig.2~c in the main text, as well as a function of the  angle $\alpha$ for fixed system size, see Fig.~\ref{fig:fig_trm}~c. 

\section{Thermalization and level spacing statistics}
In the main text, we implicitly assume eventual equilibration of the interface initial condition to a homogeneous spin state. To support this claim, which is not trivial and still under debate, we provide an analysis of the level spacing statistics of the exact eigenenergy spectrum $E_i$ for a 4 by 5 lattice with PBC in x-direction and OBC in y-direction, analogous to what has been done in the main text for larger system sizes.
The level spacing statistics provides a way to determine wether a system is chaotic or integrable, based on results from random matrix theory. An integrable system is characterized by its level spacing statistics following a Poisson distribution, while for a chaotic system it follows the one of a Gaussian orthogonal ensemble (GOE).
In order to evaluate the level spacing statistics reliably, one has to first unfold the spectrum, i.e. separate the level spacing from overall changes in the energy scale. This is done via the relation 
\begin{equation}
    s_n = (E_{n+1} - E_n) \frac{\partial n(E)}{\partial E}|_{E=E_n},
\end{equation} 
where $n(E)$ is the staircase or cumulative spectral function, returning the number of levels below or at the energy $E$~\cite{Abul_Magd_2014}.

In Figure~\ref{fig:fig_level spacing}~a) we show results of the unfolded spectrum for two different transverse fields $g/J=0.15,0.75$. The level spacing of the larger field clearly resembles the one of a GOE, indicating a chaotic system, which is true also for the smaller transverse field, even though the peak of the distribution starts to travel towards zero level spacing. To summarize, neither of the cases follows the level spacing distribution expected from an integrable system.

The mean over the distribution of ratios 
\begin{equation}
    \tilde{r}_n = \frac{\min(d_n,d_{n-1})}{\max(d_n,d_{n-1})},
\end{equation}
with $d_n=E_n-E_{n-1}$, assumes the value $\langle \tilde{r}_n \rangle_\mathrm{Poisson} \approx 0.38629$ or $\langle \tilde{r}_n \rangle_\mathrm{GOE} \approx 0.60266$ if the underlying distribution is of Poissonian or GOE type, respectively~\cite{Atas_2013}.
Figure~\ref{fig:fig_level spacing}~b) presents this quantity as a function of the transverse field, showing that in the considered system of finite size the underlying level spacing statistics closely follows the chaotic one, with minimal deviations towards smaller transverse fields. Hence, we expect eventual thermalization of the system in the non-equilibrium dynamics considered in the main text for all system sizes considered.

\begin{figure}[h!]
    \centering
    \includegraphics[width=0.7\linewidth]{figures/supplemental_material/sup_figure_levelspacing.pdf}
    \caption{a) Two examples of the unfolded level spacings at different $g/J$. The one corresponding to $g/J=0.75$ clearly agrees with the one predicted from random matrix theory for chaotic systems, while for the smaller $g/J=0.15$ it starts exhibiting features from both, i.e. the mean level spacing moves towards zero -- however the characteristic peak at $s_n=0$ is still missing, distinguishing it from a pure Poisson statistics. b) Mean level-spacing distribution ratio of a 4 by 5 system for different transverse fields. Across all transverse fields considered, the ratio corresponds to a GOE distribution and hence an chaotic system with slight deviations from that expected value at the smallest fields.} \label{fig:fig_level spacing} 
\end{figure}
\section{Quantum Monte Carlo}
The data points in the phase diagram of Fig. 2d) are obtained using the QMC \emph{loop}-algorithm~\cite{Evertz_1993,Evertz_2003} provided in the ALPS package~\cite{Bauer_2011} for the critical temperatures (blue line) and using the QMC package written in Rust~\cite{Hearth2024} for the effective temperature of the initial state (red line).

\begin{figure}[h!]
    \centering
    \includegraphics[width=0.7\linewidth]{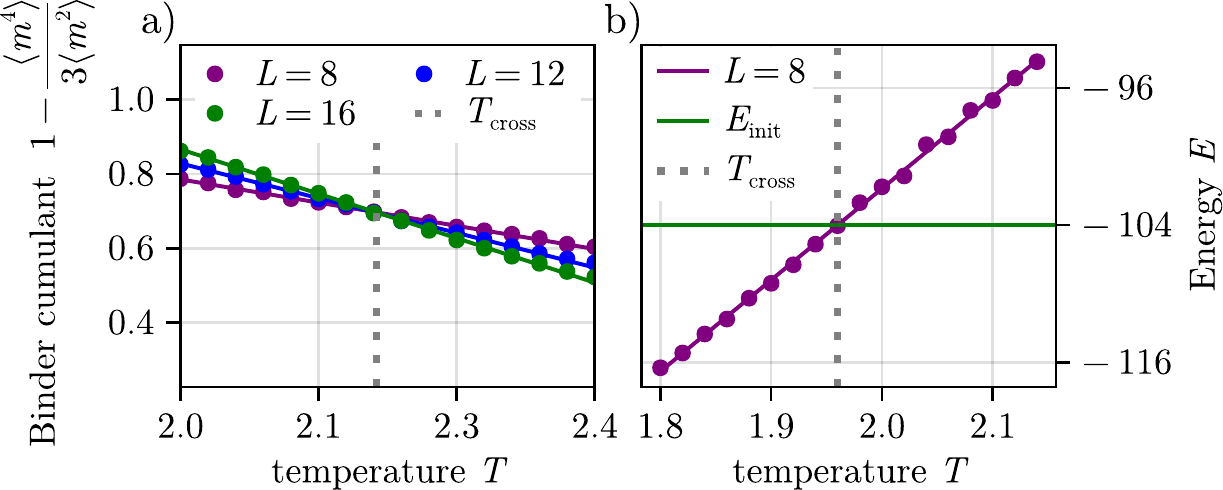}
    \caption{a) Binder cumulant of the magnetization on a 2D square lattice for three different lattice dimensions. Shown is an example for a transverse field $g/J=1$. The crossing point of the three lines marks the critical temperature of the system at said transverse field. b) This plot shows how an effective temperature can be associated to a non-equilibrium state. For that, we find the temperature of an equilibrium state whose energy coincides with the energy of the state we are interested in. This is shown again on the example of $g/J=1$ for an $8\times8$ lattice.} \label{fig:fig_qmc} 
\end{figure}

The points of the blue line are obtained by finding the crossing points of the binder cumulant of the magnetization as a function of temperature for lattice sizes $L=8,12,16$, see Fig.~\ref{fig:fig_qmc}~a. 

The points of the red line correspond to the temperature associated with the initial domain wall state on an $8 \times 8$ lattice at a transverse field $g$, i.e. the temperature $T_\mathrm{cross}$ s.t. the energy $E=\mathrm{Tr}\ H \rho_\mathrm{Gibbs}$ of the Gibbs state $\rho_\mathrm{Gibbs}=\exp{(-H / T_\mathrm{cross})}$ corresponds to the energy of the initial state $E_\mathrm{init}$, see Fig.~\ref{fig:fig_qmc}~b. 
Note, that in order to simulate the conditions for a domain wall state and get the correct effective temperature, we employ periodic boundary conditions on one edge and anti-periodic boundary conditions on the other within the QMC simulation.

Data points are obtained using $5\cdot10^4$/$1\cdot10^5$ samples for the critical/actual temperatures, after $1\cdot10^4$ thermalization steps are performed. 
These values where chosen such that the estimated error on all the observables considered here lies below $1\%$.

%% file: main.bbl
\begin{thebibliography}{64}%
\makeatletter
\providecommand \@ifxundefined [1]{%
 \@ifx{#1\undefined}
}%
\providecommand \@ifnum [1]{%
 \ifnum #1\expandafter \@firstoftwo
 \else \expandafter \@secondoftwo
 \fi
}%
\providecommand \@ifx [1]{%
 \ifx #1\expandafter \@firstoftwo
 \else \expandafter \@secondoftwo
 \fi
}%
\providecommand \natexlab [1]{#1}%
\providecommand \enquote  [1]{``#1''}%
\providecommand \bibnamefont  [1]{#1}%
\providecommand \bibfnamefont [1]{#1}%
\providecommand \citenamefont [1]{#1}%
\providecommand \href@noop [0]{\@secondoftwo}%
\providecommand \href [0]{\begingroup \@sanitize@url \@href}%
\providecommand \@href[1]{\@@startlink{#1}\@@href}%
\providecommand \@@href[1]{\endgroup#1\@@endlink}%
\providecommand \@sanitize@url [0]{\catcode `\\12\catcode `\$12\catcode
  `\&12\catcode `\#12\catcode `\^12\catcode `\_12\catcode `\%12\relax}%
\providecommand \@@startlink[1]{}%
\providecommand \@@endlink[0]{}%
\providecommand \url  [0]{\begingroup\@sanitize@url \@url }%
\providecommand \@url [1]{\endgroup\@href {#1}{\urlprefix }}%
\providecommand \urlprefix  [0]{URL }%
\providecommand \Eprint [0]{\href }%
\providecommand \doibase [0]{https://doi.org/}%
\providecommand \selectlanguage [0]{\@gobble}%
\providecommand \bibinfo  [0]{\@secondoftwo}%
\providecommand \bibfield  [0]{\@secondoftwo}%
\providecommand \translation [1]{[#1]}%
\providecommand \BibitemOpen [0]{}%
\providecommand \bibitemStop [0]{}%
\providecommand \bibitemNoStop [0]{.\EOS\space}%
\providecommand \EOS [0]{\spacefactor3000\relax}%
\providecommand \BibitemShut  [1]{\csname bibitem#1\endcsname}%
\let\auto@bib@innerbib\@empty
\bibitem [{\citenamefont {Burton}\ \emph {et~al.}(1951)\citenamefont {Burton},
  \citenamefont {Cabrera},\ and\ \citenamefont {Frank}}]{Burton1951}%
  \BibitemOpen
  \bibfield  {author} {\bibinfo {author} {\bibfnamefont {W.~K.}\ \bibnamefont
  {Burton}}, \bibinfo {author} {\bibfnamefont {N.}~\bibnamefont {Cabrera}},\
  and\ \bibinfo {author} {\bibfnamefont {F.~C.}\ \bibnamefont {Frank}},\
  }\bibfield  {title} {\bibinfo {title} {The growth of crystals and the
  equilibrium structure of their surfaces},\ }\href
  {https://doi.org/10.1098/rsta.1951.0006} {\bibfield  {journal} {\bibinfo
  {journal} {Philosophical Transactions of the Royal Society of London. Series
  A, Mathematical and Physical Sciences}\ }\textbf {\bibinfo {volume} {243}},\
  \bibinfo {pages} {299} (\bibinfo {year} {1951})},\ \bibinfo {note}
  {publisher: Royal Society}\BibitemShut {NoStop}%
\bibitem [{\citenamefont {Dobrushin}(1973)}]{Dobrushin_1973}%
  \BibitemOpen
  \bibfield  {author} {\bibinfo {author} {\bibfnamefont {R.~L.}\ \bibnamefont
  {Dobrushin}},\ }\bibfield  {title} {\bibinfo {title} {{Gibbs State Describing
  Coexistence of Phases for a Three-Dimensional Ising Model}},\ }\href
  {https://api.semanticscholar.org/CorpusID:120239349} {\bibfield  {journal}
  {\bibinfo  {journal} {Theory of Probability and Its Applications}\ }\textbf
  {\bibinfo {volume} {17}},\ \bibinfo {pages} {582} (\bibinfo {year}
  {1973})}\BibitemShut {NoStop}%
\bibitem [{\citenamefont {Weeks}\ \emph {et~al.}(1973)\citenamefont {Weeks},
  \citenamefont {Gilmer},\ and\ \citenamefont {Leamy}}]{Weeks1973}%
  \BibitemOpen
  \bibfield  {author} {\bibinfo {author} {\bibfnamefont {J.~D.}\ \bibnamefont
  {Weeks}}, \bibinfo {author} {\bibfnamefont {G.~H.}\ \bibnamefont {Gilmer}},\
  and\ \bibinfo {author} {\bibfnamefont {H.~J.}\ \bibnamefont {Leamy}},\
  }\bibfield  {title} {\bibinfo {title} {{Structural Transition in the
  Ising-Model Interface}},\ }\href {https://doi.org/10.1103/PhysRevLett.31.549}
  {\bibfield  {journal} {\bibinfo  {journal} {Phys. Rev. Lett.}\ }\textbf
  {\bibinfo {volume} {31}},\ \bibinfo {pages} {549} (\bibinfo {year}
  {1973})}\BibitemShut {NoStop}%
\bibitem [{\citenamefont {Chui}\ and\ \citenamefont {Weeks}(1976)}]{Chui_76}%
  \BibitemOpen
  \bibfield  {author} {\bibinfo {author} {\bibfnamefont {S.~T.}\ \bibnamefont
  {Chui}}\ and\ \bibinfo {author} {\bibfnamefont {J.~D.}\ \bibnamefont
  {Weeks}},\ }\bibfield  {title} {\bibinfo {title} {Phase transition in the
  two-dimensional coulomb gas, and the interfacial roughening transition},\
  }\href {https://doi.org/10.1103/PhysRevB.14.4978} {\bibfield  {journal}
  {\bibinfo  {journal} {Phys. Rev. B}\ }\textbf {\bibinfo {volume} {14}},\
  \bibinfo {pages} {4978} (\bibinfo {year} {1976})}\BibitemShut {NoStop}%
\bibitem [{\citenamefont {van Beijeren}(1977)}]{Beijeren_1977}%
  \BibitemOpen
  \bibfield  {author} {\bibinfo {author} {\bibfnamefont {H.}~\bibnamefont {van
  Beijeren}},\ }\bibfield  {title} {\bibinfo {title} {Exactly {Solvable}
  {Model} for the {Roughening} {Transition} of a {Crystal} {Surface}},\ }\href
  {https://doi.org/10.1103/PhysRevLett.38.993} {\bibfield  {journal} {\bibinfo
  {journal} {Physical Review Letters}\ }\textbf {\bibinfo {volume} {38}},\
  \bibinfo {pages} {993} (\bibinfo {year} {1977})},\ \bibinfo {note}
  {publisher: American Physical Society}\BibitemShut {NoStop}%
\bibitem [{\citenamefont {Hasenbusch}\ \emph {et~al.}(1996)\citenamefont
  {Hasenbusch}, \citenamefont {Meyer},\ and\ \citenamefont
  {Pütz}}]{Hasenbusch_1996}%
  \BibitemOpen
  \bibfield  {author} {\bibinfo {author} {\bibfnamefont {M.}~\bibnamefont
  {Hasenbusch}}, \bibinfo {author} {\bibfnamefont {S.}~\bibnamefont {Meyer}},\
  and\ \bibinfo {author} {\bibfnamefont {M.}~\bibnamefont {Pütz}},\ }\bibfield
   {title} {\bibinfo {title} {{The roughening transition of the
  three-dimensional Ising interface: A Monte Carlo study}},\ }\href
  {https://doi.org/10.1007/bf02174211} {\bibfield  {journal} {\bibinfo
  {journal} {Journal of Statistical Physics}\ }\textbf {\bibinfo {volume}
  {85}},\ \bibinfo {pages} {383–401} (\bibinfo {year} {1996})}\BibitemShut
  {NoStop}%
\bibitem [{\citenamefont {Balibar}\ \emph {et~al.}(2005)\citenamefont
  {Balibar}, \citenamefont {Alles},\ and\ \citenamefont
  {Parshin}}]{Balibar_2005}%
  \BibitemOpen
  \bibfield  {author} {\bibinfo {author} {\bibfnamefont {S.}~\bibnamefont
  {Balibar}}, \bibinfo {author} {\bibfnamefont {H.}~\bibnamefont {Alles}},\
  and\ \bibinfo {author} {\bibfnamefont {A.~Y.}\ \bibnamefont {Parshin}},\
  }\bibfield  {title} {\bibinfo {title} {The surface of helium crystals},\
  }\href {https://api.semanticscholar.org/CorpusID:56030886} {\bibfield
  {journal} {\bibinfo  {journal} {Reviews of Modern Physics}\ }\textbf
  {\bibinfo {volume} {77}},\ \bibinfo {pages} {317} (\bibinfo {year}
  {2005})}\BibitemShut {NoStop}%
\bibitem [{\citenamefont {Fisher}\ and\ \citenamefont
  {Weeks}(1983)}]{Fisher1983}%
  \BibitemOpen
  \bibfield  {author} {\bibinfo {author} {\bibfnamefont {D.~S.}\ \bibnamefont
  {Fisher}}\ and\ \bibinfo {author} {\bibfnamefont {J.~D.}\ \bibnamefont
  {Weeks}},\ }\bibfield  {title} {\bibinfo {title} {Shape of {Crystals} at
  {Low} {Temperatures}: {Absence} of {Quantum} {Roughening}},\ }\href
  {https://doi.org/10.1103/PhysRevLett.50.1077} {\bibfield  {journal} {\bibinfo
   {journal} {Physical Review Letters}\ }\textbf {\bibinfo {volume} {50}},\
  \bibinfo {pages} {1077} (\bibinfo {year} {1983})},\ \bibinfo {note}
  {publisher: American Physical Society}\BibitemShut {NoStop}%
\bibitem [{\citenamefont {{Fradkin}}(1983)}]{Fradkin1983}%
  \BibitemOpen
  \bibfield  {author} {\bibinfo {author} {\bibfnamefont {E.}~\bibnamefont
  {{Fradkin}}},\ }\bibfield  {title} {\bibinfo {title} {{Roughening transition
  in quantum interfaces}},\ }\href {https://doi.org/10.1103/PhysRevB.28.5338}
  {\bibfield  {journal} {\bibinfo  {journal} {\prb}\ }\textbf {\bibinfo
  {volume} {28}},\ \bibinfo {pages} {5338} (\bibinfo {year}
  {1983})}\BibitemShut {NoStop}%
\bibitem [{\citenamefont {{L{\"u}scher}}(1981)}]{Luescher_1981}%
  \BibitemOpen
  \bibfield  {author} {\bibinfo {author} {\bibfnamefont {M.}~\bibnamefont
  {{L{\"u}scher}}},\ }\bibfield  {title} {\bibinfo {title} {{Symmetry-breaking
  aspects of the roughening transition in gauge theories}},\ }\href
  {https://doi.org/10.1016/0550-3213(81)90423-5} {\bibfield  {journal}
  {\bibinfo  {journal} {Nuclear Physics B}\ }\textbf {\bibinfo {volume}
  {180}},\ \bibinfo {pages} {317} (\bibinfo {year} {1981})}\BibitemShut
  {NoStop}%
\bibitem [{\citenamefont {Cochran~\textit{et al.}}(2024)}]{Cochran2024}%
  \BibitemOpen
  \bibfield  {author} {\bibinfo {author} {\bibfnamefont {T.~A.}\ \bibnamefont
  {Cochran~\textit{et al.}}},\ }\bibfield  {title} {\bibinfo {title}
  {Visualizing {Dynamics} of {Charges} and {Strings} in (2+1){D} {Lattice}
  {Gauge} {Theories}},\ }\bibfield  {journal} {\bibinfo  {journal}
  {arXiv:2409.17142}\ }\href {https://doi.org/10.48550/arXiv.2409.17142}
  {10.48550/arXiv.2409.17142} (\bibinfo {year} {2024})\BibitemShut {NoStop}%
\bibitem [{\citenamefont {Kloss}\ \emph {et~al.}(2020)\citenamefont {Kloss},
  \citenamefont {Reichman},\ and\ \citenamefont {Lev}}]{Kloss_2020}%
  \BibitemOpen
  \bibfield  {author} {\bibinfo {author} {\bibfnamefont {B.}~\bibnamefont
  {Kloss}}, \bibinfo {author} {\bibfnamefont {D.~R.}\ \bibnamefont
  {Reichman}},\ and\ \bibinfo {author} {\bibfnamefont {Y.~B.}\ \bibnamefont
  {Lev}},\ }\bibfield  {title} {\bibinfo {title} {{Studying dynamics in
  two-dimensional quantum lattices using tree tensor network states}},\ }\href
  {https://doi.org/10.21468/SciPostPhys.9.5.070} {\bibfield  {journal}
  {\bibinfo  {journal} {SciPost Phys.}\ }\textbf {\bibinfo {volume} {9}},\
  \bibinfo {pages} {070} (\bibinfo {year} {2020})}\BibitemShut {NoStop}%
\bibitem [{\citenamefont {Pavešić}\ \emph {et~al.}(2024)\citenamefont
  {Pavešić}, \citenamefont {Jaschke},\ and\ \citenamefont
  {Montangero}}]{Pavesic_2024}%
  \BibitemOpen
  \bibfield  {author} {\bibinfo {author} {\bibfnamefont {L.}~\bibnamefont
  {Pavešić}}, \bibinfo {author} {\bibfnamefont {D.}~\bibnamefont {Jaschke}},\
  and\ \bibinfo {author} {\bibfnamefont {S.}~\bibnamefont {Montangero}},\
  }\href {https://arxiv.org/abs/2406.11979} {\bibinfo {title} {{Constrained
  dynamics and confinement in the two-dimensional quantum Ising model}}}
  (\bibinfo {year} {2024}),\ \Eprint {https://arxiv.org/abs/2406.11979}
  {arXiv:2406.11979} \BibitemShut {NoStop}%
\bibitem [{\citenamefont {Krinitsin}\ \emph
  {et~al.}(2025{\natexlab{a}})\citenamefont {Krinitsin}, \citenamefont
  {Tausendpfund}, \citenamefont {Heyl}, \citenamefont {Rizzi},\ and\
  \citenamefont {Schmitt}}]{krinitsin2025_quantumising}%
  \BibitemOpen
  \bibfield  {author} {\bibinfo {author} {\bibfnamefont {W.}~\bibnamefont
  {Krinitsin}}, \bibinfo {author} {\bibfnamefont {N.}~\bibnamefont
  {Tausendpfund}}, \bibinfo {author} {\bibfnamefont {M.}~\bibnamefont {Heyl}},
  \bibinfo {author} {\bibfnamefont {M.}~\bibnamefont {Rizzi}},\ and\ \bibinfo
  {author} {\bibfnamefont {M.}~\bibnamefont {Schmitt}},\ }\href
  {https://arxiv.org/abs/2505.07612} {\bibinfo {title} {Time evolution of the
  quantum ising model in two dimensions using tree tensor networks}} (\bibinfo
  {year} {2025}{\natexlab{a}}),\ \Eprint {https://arxiv.org/abs/2505.07612}
  {arXiv:2505.07612 [quant-ph]} \BibitemShut {NoStop}%
\bibitem [{\citenamefont {Moeckel}\ and\ \citenamefont
  {Kehrein}(2008)}]{Kehrein2008}%
  \BibitemOpen
  \bibfield  {author} {\bibinfo {author} {\bibfnamefont {M.}~\bibnamefont
  {Moeckel}}\ and\ \bibinfo {author} {\bibfnamefont {S.}~\bibnamefont
  {Kehrein}},\ }\bibfield  {title} {\bibinfo {title} {Interaction quench in the
  hubbard model},\ }\href {https://doi.org/10.1103/PhysRevLett.100.175702}
  {\bibfield  {journal} {\bibinfo  {journal} {Phys. Rev. Lett.}\ }\textbf
  {\bibinfo {volume} {100}},\ \bibinfo {pages} {175702} (\bibinfo {year}
  {2008})}\BibitemShut {NoStop}%
\bibitem [{\citenamefont {Moudgalya}\ \emph {et~al.}(2022)\citenamefont
  {Moudgalya}, \citenamefont {Bernevig},\ and\ \citenamefont
  {Regnault}}]{Moudgalya2022}%
  \BibitemOpen
  \bibfield  {author} {\bibinfo {author} {\bibfnamefont {S.}~\bibnamefont
  {Moudgalya}}, \bibinfo {author} {\bibfnamefont {B.~A.}\ \bibnamefont
  {Bernevig}},\ and\ \bibinfo {author} {\bibfnamefont {N.}~\bibnamefont
  {Regnault}},\ }\bibfield  {title} {\bibinfo {title} {{Quantum many-body scars
  and Hilbert space fragmentation: a review of exact results}},\ }\href
  {https://doi.org/10.1088/1361-6633/ac73a0} {\bibfield  {journal} {\bibinfo
  {journal} {Reports on Progress in Physics}\ }\textbf {\bibinfo {volume}
  {85}},\ \bibinfo {pages} {086501} (\bibinfo {year} {2022})}\BibitemShut
  {NoStop}%
\bibitem [{\citenamefont {Greiner}\ \emph {et~al.}(2002)\citenamefont
  {Greiner}, \citenamefont {Mandel}, \citenamefont {Hänsch},\ and\
  \citenamefont {Bloch}}]{Greiner_2002}%
  \BibitemOpen
  \bibfield  {author} {\bibinfo {author} {\bibfnamefont {M.}~\bibnamefont
  {Greiner}}, \bibinfo {author} {\bibfnamefont {O.}~\bibnamefont {Mandel}},
  \bibinfo {author} {\bibfnamefont {T.~W.}\ \bibnamefont {Hänsch}},\ and\
  \bibinfo {author} {\bibfnamefont {I.}~\bibnamefont {Bloch}},\ }\bibfield
  {title} {\bibinfo {title} {{Collapse and revival of the matter wave field of
  a Bose–Einstein condensate}},\ }\href {https://doi.org/10.1038/nature00968}
  {\bibfield  {journal} {\bibinfo  {journal} {Nature}\ }\textbf {\bibinfo
  {volume} {419}},\ \bibinfo {pages} {51–54} (\bibinfo {year}
  {2002})}\BibitemShut {NoStop}%
\bibitem [{\citenamefont {Georgescu}\ \emph {et~al.}(2014)\citenamefont
  {Georgescu}, \citenamefont {Ashhab},\ and\ \citenamefont
  {Nori}}]{Georgescu_2014}%
  \BibitemOpen
  \bibfield  {author} {\bibinfo {author} {\bibfnamefont {I.~M.}\ \bibnamefont
  {Georgescu}}, \bibinfo {author} {\bibfnamefont {S.}~\bibnamefont {Ashhab}},\
  and\ \bibinfo {author} {\bibfnamefont {F.}~\bibnamefont {Nori}},\ }\bibfield
  {title} {\bibinfo {title} {Quantum simulation},\ }\href
  {https://doi.org/10.1103/RevModPhys.86.153} {\bibfield  {journal} {\bibinfo
  {journal} {Rev. Mod. Phys.}\ }\textbf {\bibinfo {volume} {86}},\ \bibinfo
  {pages} {153} (\bibinfo {year} {2014})}\BibitemShut {NoStop}%
\bibitem [{\citenamefont {Kinoshita}\ \emph {et~al.}(2006)\citenamefont
  {Kinoshita}, \citenamefont {Wenger},\ and\ \citenamefont
  {Weiss}}]{Kinoshita_2006}%
  \BibitemOpen
  \bibfield  {author} {\bibinfo {author} {\bibfnamefont {T.}~\bibnamefont
  {Kinoshita}}, \bibinfo {author} {\bibfnamefont {T.}~\bibnamefont {Wenger}},\
  and\ \bibinfo {author} {\bibfnamefont {D.~S.}\ \bibnamefont {Weiss}},\
  }\bibfield  {title} {\bibinfo {title} {{A quantum Newton's cradle}},\ }\href
  {https://doi.org/10.1038/nature04693} {\bibfield  {journal} {\bibinfo
  {journal} {Nature}\ }\textbf {\bibinfo {volume} {440}},\ \bibinfo {pages}
  {900} (\bibinfo {year} {2006})}\BibitemShut {NoStop}%
\bibitem [{\citenamefont {Martinez}\ \emph {et~al.}(2016)\citenamefont
  {Martinez}, \citenamefont {Muschik}, \citenamefont {Schindler}, \citenamefont
  {Nigg}, \citenamefont {Erhard}, \citenamefont {Heyl}, \citenamefont {Hauke},
  \citenamefont {Dalmonte}, \citenamefont {Monz}, \citenamefont {Zoller},\ and\
  \citenamefont {Blatt}}]{Martinez_2016}%
  \BibitemOpen
  \bibfield  {author} {\bibinfo {author} {\bibfnamefont {E.~A.}\ \bibnamefont
  {Martinez}}, \bibinfo {author} {\bibfnamefont {C.~A.}\ \bibnamefont
  {Muschik}}, \bibinfo {author} {\bibfnamefont {P.}~\bibnamefont {Schindler}},
  \bibinfo {author} {\bibfnamefont {D.}~\bibnamefont {Nigg}}, \bibinfo {author}
  {\bibfnamefont {A.}~\bibnamefont {Erhard}}, \bibinfo {author} {\bibfnamefont
  {M.}~\bibnamefont {Heyl}}, \bibinfo {author} {\bibfnamefont {P.}~\bibnamefont
  {Hauke}}, \bibinfo {author} {\bibfnamefont {M.}~\bibnamefont {Dalmonte}},
  \bibinfo {author} {\bibfnamefont {T.}~\bibnamefont {Monz}}, \bibinfo {author}
  {\bibfnamefont {P.}~\bibnamefont {Zoller}},\ and\ \bibinfo {author}
  {\bibfnamefont {R.}~\bibnamefont {Blatt}},\ }\bibfield  {title} {\bibinfo
  {title} {Real-time dynamics of lattice gauge theories with a few-qubit
  quantum computer},\ }\href {https://doi.org/10.1038/nature18318} {\bibfield
  {journal} {\bibinfo  {journal} {Nature}\ }\textbf {\bibinfo {volume} {534}},\
  \bibinfo {pages} {516–519} (\bibinfo {year} {2016})}\BibitemShut {NoStop}%
\bibitem [{\citenamefont {Choi}\ \emph {et~al.}(2016)\citenamefont {Choi},
  \citenamefont {Hild}, \citenamefont {Zeiher}, \citenamefont {Schauß},
  \citenamefont {Rubio-Abadal}, \citenamefont {Yefsah}, \citenamefont
  {Khemani}, \citenamefont {Huse}, \citenamefont {Bloch},\ and\ \citenamefont
  {Gross}}]{Jae-yoon_2016}%
  \BibitemOpen
  \bibfield  {author} {\bibinfo {author} {\bibfnamefont {J.-Y.}\ \bibnamefont
  {Choi}}, \bibinfo {author} {\bibfnamefont {S.}~\bibnamefont {Hild}}, \bibinfo
  {author} {\bibfnamefont {J.}~\bibnamefont {Zeiher}}, \bibinfo {author}
  {\bibfnamefont {P.}~\bibnamefont {Schauß}}, \bibinfo {author} {\bibfnamefont
  {A.}~\bibnamefont {Rubio-Abadal}}, \bibinfo {author} {\bibfnamefont
  {T.}~\bibnamefont {Yefsah}}, \bibinfo {author} {\bibfnamefont
  {V.}~\bibnamefont {Khemani}}, \bibinfo {author} {\bibfnamefont {D.~A.}\
  \bibnamefont {Huse}}, \bibinfo {author} {\bibfnamefont {I.}~\bibnamefont
  {Bloch}},\ and\ \bibinfo {author} {\bibfnamefont {C.}~\bibnamefont {Gross}},\
  }\bibfield  {title} {\bibinfo {title} {Exploring the many-body localization
  transition in two dimensions},\ }\href
  {https://doi.org/10.1126/science.aaf8834} {\bibfield  {journal} {\bibinfo
  {journal} {Science}\ }\textbf {\bibinfo {volume} {352}},\ \bibinfo {pages}
  {1547} (\bibinfo {year} {2016})},\ \Eprint
  {https://arxiv.org/abs/https://www.science.org/doi/pdf/10.1126/science.aaf8834}
  {https://www.science.org/doi/pdf/10.1126/science.aaf8834} \BibitemShut
  {NoStop}%
\bibitem [{\citenamefont {Gross}\ and\ \citenamefont
  {Bloch}(2017)}]{Gross_2017}%
  \BibitemOpen
  \bibfield  {author} {\bibinfo {author} {\bibfnamefont {C.}~\bibnamefont
  {Gross}}\ and\ \bibinfo {author} {\bibfnamefont {I.}~\bibnamefont {Bloch}},\
  }\bibfield  {title} {\bibinfo {title} {Quantum simulations with ultracold
  atoms in optical lattices},\ }\href {https://doi.org/10.1126/science.aal3837}
  {\bibfield  {journal} {\bibinfo  {journal} {Science}\ }\textbf {\bibinfo
  {volume} {357}},\ \bibinfo {pages} {995} (\bibinfo {year} {2017})},\ \Eprint
  {https://arxiv.org/abs/https://www.science.org/doi/pdf/10.1126/science.aal3837}
  {https://www.science.org/doi/pdf/10.1126/science.aal3837} \BibitemShut
  {NoStop}%
\bibitem [{\citenamefont {Jurcevic}\ \emph {et~al.}(2017)\citenamefont
  {Jurcevic}, \citenamefont {Shen}, \citenamefont {Hauke}, \citenamefont
  {Maier}, \citenamefont {Brydges}, \citenamefont {Hempel}, \citenamefont
  {Lanyon}, \citenamefont {Heyl}, \citenamefont {Blatt},\ and\ \citenamefont
  {Roos}}]{Jurcevic_2017}%
  \BibitemOpen
  \bibfield  {author} {\bibinfo {author} {\bibfnamefont {P.}~\bibnamefont
  {Jurcevic}}, \bibinfo {author} {\bibfnamefont {H.}~\bibnamefont {Shen}},
  \bibinfo {author} {\bibfnamefont {P.}~\bibnamefont {Hauke}}, \bibinfo
  {author} {\bibfnamefont {C.}~\bibnamefont {Maier}}, \bibinfo {author}
  {\bibfnamefont {T.}~\bibnamefont {Brydges}}, \bibinfo {author} {\bibfnamefont
  {C.}~\bibnamefont {Hempel}}, \bibinfo {author} {\bibfnamefont {B.~P.}\
  \bibnamefont {Lanyon}}, \bibinfo {author} {\bibfnamefont {M.}~\bibnamefont
  {Heyl}}, \bibinfo {author} {\bibfnamefont {R.}~\bibnamefont {Blatt}},\ and\
  \bibinfo {author} {\bibfnamefont {C.~F.}\ \bibnamefont {Roos}},\ }\bibfield
  {title} {\bibinfo {title} {Direct observation of dynamical quantum phase
  transitions in an interacting many-body system},\ }\href
  {https://doi.org/10.1103/PhysRevLett.119.080501} {\bibfield  {journal}
  {\bibinfo  {journal} {Phys. Rev. Lett.}\ }\textbf {\bibinfo {volume} {119}},\
  \bibinfo {pages} {080501} (\bibinfo {year} {2017})}\BibitemShut {NoStop}%
\bibitem [{\citenamefont {Bernien}\ \emph
  {et~al.}(2017{\natexlab{a}})\citenamefont {Bernien}, \citenamefont
  {Schwartz}, \citenamefont {Keesling}, \citenamefont {Levine}, \citenamefont
  {Omran}, \citenamefont {Pichler}, \citenamefont {Choi}, \citenamefont
  {Zibrov}, \citenamefont {Endres}, \citenamefont {Greiner}, \citenamefont
  {Vuletić},\ and\ \citenamefont {Lukin}}]{Bernien_2017}%
  \BibitemOpen
  \bibfield  {author} {\bibinfo {author} {\bibfnamefont {H.}~\bibnamefont
  {Bernien}}, \bibinfo {author} {\bibfnamefont {S.}~\bibnamefont {Schwartz}},
  \bibinfo {author} {\bibfnamefont {A.}~\bibnamefont {Keesling}}, \bibinfo
  {author} {\bibfnamefont {H.}~\bibnamefont {Levine}}, \bibinfo {author}
  {\bibfnamefont {A.}~\bibnamefont {Omran}}, \bibinfo {author} {\bibfnamefont
  {H.}~\bibnamefont {Pichler}}, \bibinfo {author} {\bibfnamefont
  {S.}~\bibnamefont {Choi}}, \bibinfo {author} {\bibfnamefont {A.~S.}\
  \bibnamefont {Zibrov}}, \bibinfo {author} {\bibfnamefont {M.}~\bibnamefont
  {Endres}}, \bibinfo {author} {\bibfnamefont {M.}~\bibnamefont {Greiner}},
  \bibinfo {author} {\bibfnamefont {V.}~\bibnamefont {Vuletić}},\ and\
  \bibinfo {author} {\bibfnamefont {M.~D.}\ \bibnamefont {Lukin}},\ }\bibfield
  {title} {\bibinfo {title} {Probing many-body dynamics on a 51-atom quantum
  simulator},\ }\href {https://doi.org/10.1038/nature24622} {\bibfield
  {journal} {\bibinfo  {journal} {Nature}\ }\textbf {\bibinfo {volume} {551}},\
  \bibinfo {pages} {579–584} (\bibinfo {year}
  {2017}{\natexlab{a}})}\BibitemShut {NoStop}%
\bibitem [{\citenamefont {Zhang}\ \emph {et~al.}(2017)\citenamefont {Zhang},
  \citenamefont {Pagano}, \citenamefont {Hess}, \citenamefont {Kyprianidis},
  \citenamefont {Becker}, \citenamefont {Kaplan}, \citenamefont {Gorshkov},
  \citenamefont {Gong},\ and\ \citenamefont {Monroe}}]{Zhang_2017}%
  \BibitemOpen
  \bibfield  {author} {\bibinfo {author} {\bibfnamefont {J.}~\bibnamefont
  {Zhang}}, \bibinfo {author} {\bibfnamefont {G.}~\bibnamefont {Pagano}},
  \bibinfo {author} {\bibfnamefont {P.~W.}\ \bibnamefont {Hess}}, \bibinfo
  {author} {\bibfnamefont {A.}~\bibnamefont {Kyprianidis}}, \bibinfo {author}
  {\bibfnamefont {P.}~\bibnamefont {Becker}}, \bibinfo {author} {\bibfnamefont
  {H.}~\bibnamefont {Kaplan}}, \bibinfo {author} {\bibfnamefont {A.~V.}\
  \bibnamefont {Gorshkov}}, \bibinfo {author} {\bibfnamefont {Z.-X.}\
  \bibnamefont {Gong}},\ and\ \bibinfo {author} {\bibfnamefont
  {C.}~\bibnamefont {Monroe}},\ }\bibfield  {title} {\bibinfo {title}
  {Observation of a many-body dynamical phase transition with a 53-qubit
  quantum simulator},\ }\href {https://doi.org/10.1038/nature24654} {\bibfield
  {journal} {\bibinfo  {journal} {Nature}\ }\textbf {\bibinfo {volume} {551}},\
  \bibinfo {pages} {601–604} (\bibinfo {year} {2017})}\BibitemShut {NoStop}%
\bibitem [{\citenamefont {Gärttner}\ \emph {et~al.}(2017)\citenamefont
  {Gärttner}, \citenamefont {Bohnet}, \citenamefont {Safavi-Naini},
  \citenamefont {Wall}, \citenamefont {Bollinger},\ and\ \citenamefont
  {Rey}}]{Gaerttner_2017}%
  \BibitemOpen
  \bibfield  {author} {\bibinfo {author} {\bibfnamefont {M.}~\bibnamefont
  {Gärttner}}, \bibinfo {author} {\bibfnamefont {J.~G.}\ \bibnamefont
  {Bohnet}}, \bibinfo {author} {\bibfnamefont {A.}~\bibnamefont
  {Safavi-Naini}}, \bibinfo {author} {\bibfnamefont {M.~L.}\ \bibnamefont
  {Wall}}, \bibinfo {author} {\bibfnamefont {J.~J.}\ \bibnamefont
  {Bollinger}},\ and\ \bibinfo {author} {\bibfnamefont {A.~M.}\ \bibnamefont
  {Rey}},\ }\bibfield  {title} {\bibinfo {title} {Measuring out-of-time-order
  correlations and multiple quantum spectra in a trapped-ion quantum magnet},\
  }\href {https://doi.org/10.1038/nphys4119} {\bibfield  {journal} {\bibinfo
  {journal} {Nature Physics}\ }\textbf {\bibinfo {volume} {13}},\ \bibinfo
  {pages} {781–786} (\bibinfo {year} {2017})}\BibitemShut {NoStop}%
\bibitem [{\citenamefont {Choi}\ \emph {et~al.}(2017)\citenamefont {Choi},
  \citenamefont {Choi}, \citenamefont {Landig}, \citenamefont {Kucsko},
  \citenamefont {Zhou}, \citenamefont {Isoya}, \citenamefont {Jelezko},
  \citenamefont {Onoda}, \citenamefont {Sumiya}, \citenamefont {Khemani},
  \citenamefont {von Keyserlingk}, \citenamefont {Yao}, \citenamefont
  {Demler},\ and\ \citenamefont {Lukin}}]{Choi_2017}%
  \BibitemOpen
  \bibfield  {author} {\bibinfo {author} {\bibfnamefont {S.}~\bibnamefont
  {Choi}}, \bibinfo {author} {\bibfnamefont {J.}~\bibnamefont {Choi}}, \bibinfo
  {author} {\bibfnamefont {R.}~\bibnamefont {Landig}}, \bibinfo {author}
  {\bibfnamefont {G.}~\bibnamefont {Kucsko}}, \bibinfo {author} {\bibfnamefont
  {H.}~\bibnamefont {Zhou}}, \bibinfo {author} {\bibfnamefont {J.}~\bibnamefont
  {Isoya}}, \bibinfo {author} {\bibfnamefont {F.}~\bibnamefont {Jelezko}},
  \bibinfo {author} {\bibfnamefont {S.}~\bibnamefont {Onoda}}, \bibinfo
  {author} {\bibfnamefont {H.}~\bibnamefont {Sumiya}}, \bibinfo {author}
  {\bibfnamefont {V.}~\bibnamefont {Khemani}}, \bibinfo {author} {\bibfnamefont
  {C.}~\bibnamefont {von Keyserlingk}}, \bibinfo {author} {\bibfnamefont
  {N.~Y.}\ \bibnamefont {Yao}}, \bibinfo {author} {\bibfnamefont
  {E.}~\bibnamefont {Demler}},\ and\ \bibinfo {author} {\bibfnamefont {M.~D.}\
  \bibnamefont {Lukin}},\ }\bibfield  {title} {\bibinfo {title} {Observation of
  discrete time-crystalline order in a disordered dipolar many-body system},\
  }\href {https://doi.org/10.1038/nature21426} {\bibfield  {journal} {\bibinfo
  {journal} {Nature}\ }\textbf {\bibinfo {volume} {543}},\ \bibinfo {pages}
  {221–225} (\bibinfo {year} {2017})}\BibitemShut {NoStop}%
\bibitem [{\citenamefont {Levine}\ \emph {et~al.}(2018)\citenamefont {Levine},
  \citenamefont {Keesling}, \citenamefont {Omran}, \citenamefont {Bernien},
  \citenamefont {Schwartz}, \citenamefont {Zibrov}, \citenamefont {Endres},
  \citenamefont {Greiner}, \citenamefont {Vuletić},\ and\ \citenamefont
  {Lukin}}]{Levine_2018}%
  \BibitemOpen
  \bibfield  {author} {\bibinfo {author} {\bibfnamefont {H.}~\bibnamefont
  {Levine}}, \bibinfo {author} {\bibfnamefont {A.}~\bibnamefont {Keesling}},
  \bibinfo {author} {\bibfnamefont {A.}~\bibnamefont {Omran}}, \bibinfo
  {author} {\bibfnamefont {H.}~\bibnamefont {Bernien}}, \bibinfo {author}
  {\bibfnamefont {S.}~\bibnamefont {Schwartz}}, \bibinfo {author}
  {\bibfnamefont {A.~S.}\ \bibnamefont {Zibrov}}, \bibinfo {author}
  {\bibfnamefont {M.}~\bibnamefont {Endres}}, \bibinfo {author} {\bibfnamefont
  {M.}~\bibnamefont {Greiner}}, \bibinfo {author} {\bibfnamefont
  {V.}~\bibnamefont {Vuletić}},\ and\ \bibinfo {author} {\bibfnamefont
  {M.~D.}\ \bibnamefont {Lukin}},\ }\bibfield  {title} {\bibinfo {title}
  {High-fidelity control and entanglement of rydberg-atom qubits},\ }\bibfield
  {journal} {\bibinfo  {journal} {Physical Review Letters}\ }\textbf {\bibinfo
  {volume} {121}},\ \href {https://doi.org/10.1103/physrevlett.121.123603}
  {10.1103/physrevlett.121.123603} (\bibinfo {year} {2018})\BibitemShut
  {NoStop}%
\bibitem [{\citenamefont {Hild}\ \emph {et~al.}(2014)\citenamefont {Hild},
  \citenamefont {Fukuhara}, \citenamefont {Schau\ss{}}, \citenamefont {Zeiher},
  \citenamefont {Knap}, \citenamefont {Demler}, \citenamefont {Bloch},\ and\
  \citenamefont {Gross}}]{Hild_2014}%
  \BibitemOpen
  \bibfield  {author} {\bibinfo {author} {\bibfnamefont {S.}~\bibnamefont
  {Hild}}, \bibinfo {author} {\bibfnamefont {T.}~\bibnamefont {Fukuhara}},
  \bibinfo {author} {\bibfnamefont {P.}~\bibnamefont {Schau\ss{}}}, \bibinfo
  {author} {\bibfnamefont {J.}~\bibnamefont {Zeiher}}, \bibinfo {author}
  {\bibfnamefont {M.}~\bibnamefont {Knap}}, \bibinfo {author} {\bibfnamefont
  {E.}~\bibnamefont {Demler}}, \bibinfo {author} {\bibfnamefont
  {I.}~\bibnamefont {Bloch}},\ and\ \bibinfo {author} {\bibfnamefont
  {C.}~\bibnamefont {Gross}},\ }\bibfield  {title} {\bibinfo {title}
  {Far-from-equilibrium spin transport in heisenberg quantum magnets},\ }\href
  {https://doi.org/10.1103/PhysRevLett.113.147205} {\bibfield  {journal}
  {\bibinfo  {journal} {Phys. Rev. Lett.}\ }\textbf {\bibinfo {volume} {113}},\
  \bibinfo {pages} {147205} (\bibinfo {year} {2014})}\BibitemShut {NoStop}%
\bibitem [{\citenamefont {Barredo}\ \emph {et~al.}(2018)\citenamefont
  {Barredo}, \citenamefont {Lienhard}, \citenamefont {de~Léséleuc},
  \citenamefont {Lahaye},\ and\ \citenamefont {Browaeys}}]{Barredo_2018}%
  \BibitemOpen
  \bibfield  {author} {\bibinfo {author} {\bibfnamefont {D.}~\bibnamefont
  {Barredo}}, \bibinfo {author} {\bibfnamefont {V.}~\bibnamefont {Lienhard}},
  \bibinfo {author} {\bibfnamefont {S.}~\bibnamefont {de~Léséleuc}}, \bibinfo
  {author} {\bibfnamefont {T.}~\bibnamefont {Lahaye}},\ and\ \bibinfo {author}
  {\bibfnamefont {A.}~\bibnamefont {Browaeys}},\ }\bibfield  {title} {\bibinfo
  {title} {Synthetic three-dimensional atomic structures assembled atom by
  atom},\ }\href {https://doi.org/10.1038/s41586-018-0450-2} {\bibfield
  {journal} {\bibinfo  {journal} {Nature}\ }\textbf {\bibinfo {volume} {561}},\
  \bibinfo {pages} {79–82} (\bibinfo {year} {2018})}\BibitemShut {NoStop}%
\bibitem [{\citenamefont {Manovitz}\ \emph {et~al.}(2024)\citenamefont
  {Manovitz}, \citenamefont {Li}, \citenamefont {Ebadi}, \citenamefont
  {Samajdar}, \citenamefont {Geim}, \citenamefont {Evered}, \citenamefont
  {Bluvstein}, \citenamefont {Zhou}, \citenamefont {Koyluoglu}, \citenamefont
  {Feldmeier}, \citenamefont {Dolgirev}, \citenamefont {Maskara}, \citenamefont
  {Kalinowski}, \citenamefont {Sachdev}, \citenamefont {Huse}, \citenamefont
  {Greiner}, \citenamefont {Vuleti{\ifmmode\acute{c}\else\'{c}\fi}},\ and\
  \citenamefont {Lukin}}]{Manovitz_2024}%
  \BibitemOpen
  \bibfield  {author} {\bibinfo {author} {\bibfnamefont {T.}~\bibnamefont
  {Manovitz}}, \bibinfo {author} {\bibfnamefont {S.~H.}\ \bibnamefont {Li}},
  \bibinfo {author} {\bibfnamefont {S.}~\bibnamefont {Ebadi}}, \bibinfo
  {author} {\bibfnamefont {R.}~\bibnamefont {Samajdar}}, \bibinfo {author}
  {\bibfnamefont {A.~A.}\ \bibnamefont {Geim}}, \bibinfo {author}
  {\bibfnamefont {S.~J.}\ \bibnamefont {Evered}}, \bibinfo {author}
  {\bibfnamefont {D.}~\bibnamefont {Bluvstein}}, \bibinfo {author}
  {\bibfnamefont {H.}~\bibnamefont {Zhou}}, \bibinfo {author} {\bibfnamefont
  {N.~U.}\ \bibnamefont {Koyluoglu}}, \bibinfo {author} {\bibfnamefont
  {J.}~\bibnamefont {Feldmeier}}, \bibinfo {author} {\bibfnamefont {P.~E.}\
  \bibnamefont {Dolgirev}}, \bibinfo {author} {\bibfnamefont {N.}~\bibnamefont
  {Maskara}}, \bibinfo {author} {\bibfnamefont {M.}~\bibnamefont {Kalinowski}},
  \bibinfo {author} {\bibfnamefont {S.}~\bibnamefont {Sachdev}}, \bibinfo
  {author} {\bibfnamefont {D.~A.}\ \bibnamefont {Huse}}, \bibinfo {author}
  {\bibfnamefont {M.}~\bibnamefont {Greiner}}, \bibinfo {author} {\bibfnamefont
  {V.}~\bibnamefont {Vuleti{\ifmmode\acute{c}\else\'{c}\fi}}},\ and\ \bibinfo
  {author} {\bibfnamefont {M.~D.}\ \bibnamefont {Lukin}},\ }\bibfield  {title}
  {\bibinfo {title} {{Quantum coarsening and collective dynamics on a
  programmable quantum simulator}},\ }\bibfield  {journal} {\bibinfo  {journal}
  {arXiv}\ }\href {https://doi.org/10.48550/arXiv.2407.03249}
  {10.48550/arXiv.2407.03249} (\bibinfo {year} {2024}),\ \Eprint
  {https://arxiv.org/abs/2407.03249} {2407.03249} \BibitemShut {NoStop}%
\bibitem [{\citenamefont {Bl\"ote}\ and\ \citenamefont
  {Deng}(2002)}]{Bloete_2002}%
  \BibitemOpen
  \bibfield  {author} {\bibinfo {author} {\bibfnamefont {H.~W.~J.}\
  \bibnamefont {Bl\"ote}}\ and\ \bibinfo {author} {\bibfnamefont
  {Y.}~\bibnamefont {Deng}},\ }\bibfield  {title} {\bibinfo {title} {{Cluster
  Monte Carlo simulation of the transverse Ising model}},\ }\href
  {https://doi.org/10.1103/PhysRevE.66.066110} {\bibfield  {journal} {\bibinfo
  {journal} {Phys. Rev. E}\ }\textbf {\bibinfo {volume} {66}},\ \bibinfo
  {pages} {066110} (\bibinfo {year} {2002})}\BibitemShut {NoStop}%
\bibitem [{\citenamefont {Balducci}\ \emph {et~al.}(2022)\citenamefont
  {Balducci}, \citenamefont {Gambassi}, \citenamefont {Lerose}, \citenamefont
  {Scardicchio},\ and\ \citenamefont {Vanoni}}]{Balducci_2022}%
  \BibitemOpen
  \bibfield  {author} {\bibinfo {author} {\bibfnamefont {F.}~\bibnamefont
  {Balducci}}, \bibinfo {author} {\bibfnamefont {A.}~\bibnamefont {Gambassi}},
  \bibinfo {author} {\bibfnamefont {A.}~\bibnamefont {Lerose}}, \bibinfo
  {author} {\bibfnamefont {A.}~\bibnamefont {Scardicchio}},\ and\ \bibinfo
  {author} {\bibfnamefont {C.}~\bibnamefont {Vanoni}},\ }\bibfield  {title}
  {\bibinfo {title} {{Localization and Melting of Interfaces in the
  Two-Dimensional Quantum Ising Model}},\ }\href
  {https://doi.org/10.1103/PhysRevLett.129.120601} {\bibfield  {journal}
  {\bibinfo  {journal} {Phys. Rev. Lett.}\ }\textbf {\bibinfo {volume} {129}},\
  \bibinfo {pages} {120601} (\bibinfo {year} {2022})}\BibitemShut {NoStop}%
\bibitem [{\citenamefont {Balducci}\ \emph {et~al.}(2023)\citenamefont
  {Balducci}, \citenamefont {Gambassi}, \citenamefont {Lerose}, \citenamefont
  {Scardicchio},\ and\ \citenamefont {Vanoni}}]{Balducci_2023}%
  \BibitemOpen
  \bibfield  {author} {\bibinfo {author} {\bibfnamefont {F.}~\bibnamefont
  {Balducci}}, \bibinfo {author} {\bibfnamefont {A.}~\bibnamefont {Gambassi}},
  \bibinfo {author} {\bibfnamefont {A.}~\bibnamefont {Lerose}}, \bibinfo
  {author} {\bibfnamefont {A.}~\bibnamefont {Scardicchio}},\ and\ \bibinfo
  {author} {\bibfnamefont {C.}~\bibnamefont {Vanoni}},\ }\bibfield  {title}
  {\bibinfo {title} {{Interface dynamics in the two-dimensional quantum Ising
  model}},\ }\href {https://doi.org/10.1103/PhysRevB.107.024306} {\bibfield
  {journal} {\bibinfo  {journal} {Phys. Rev. B}\ }\textbf {\bibinfo {volume}
  {107}},\ \bibinfo {pages} {024306} (\bibinfo {year} {2023})}\BibitemShut
  {NoStop}%
\bibitem [{Sup()}]{Supplemental}%
  \BibitemOpen
  \href@noop {} {\bibinfo {title} {{See Supplemental Material for further
  details about the tree tensor network simulations, properties of the kink
  operator, the transfer matrix analysis, and the quantum Monte Carlo
  simulations.}}}\BibitemShut {Stop}%
\bibitem [{\citenamefont {Silvi}\ \emph {et~al.}(2019)\citenamefont {Silvi},
  \citenamefont {Tschirsich}, \citenamefont {Gerster}, \citenamefont
  {Jünemann}, \citenamefont {Jaschke}, \citenamefont {Rizzi},\ and\
  \citenamefont {Montangero}}]{Silvi2019}%
  \BibitemOpen
  \bibfield  {author} {\bibinfo {author} {\bibfnamefont {P.}~\bibnamefont
  {Silvi}}, \bibinfo {author} {\bibfnamefont {F.}~\bibnamefont {Tschirsich}},
  \bibinfo {author} {\bibfnamefont {M.}~\bibnamefont {Gerster}}, \bibinfo
  {author} {\bibfnamefont {J.}~\bibnamefont {Jünemann}}, \bibinfo {author}
  {\bibfnamefont {D.}~\bibnamefont {Jaschke}}, \bibinfo {author} {\bibfnamefont
  {M.}~\bibnamefont {Rizzi}},\ and\ \bibinfo {author} {\bibfnamefont
  {S.}~\bibnamefont {Montangero}},\ }\bibfield  {title} {\bibinfo {title} {{The
  Tensor Networks Anthology: Simulation techniques for many-body quantum
  lattice systems}},\ }\href {https://doi.org/10.21468/SciPostPhysLectNotes.8}
  {\bibfield  {journal} {\bibinfo  {journal} {SciPost Phys. Lect. Notes}\ ,\
  \bibinfo {pages} {8}} (\bibinfo {year} {2019})}\BibitemShut {NoStop}%
\bibitem [{\citenamefont {Haegeman}\ \emph {et~al.}(2011)\citenamefont
  {Haegeman}, \citenamefont {Cirac}, \citenamefont {Osborne}, \citenamefont
  {Pi\ifmmode~\check{z}\else \v{z}\fi{}orn}, \citenamefont {Verschelde},\ and\
  \citenamefont {Verstraete}}]{Haegeman2011}%
  \BibitemOpen
  \bibfield  {author} {\bibinfo {author} {\bibfnamefont {J.}~\bibnamefont
  {Haegeman}}, \bibinfo {author} {\bibfnamefont {J.~I.}\ \bibnamefont {Cirac}},
  \bibinfo {author} {\bibfnamefont {T.~J.}\ \bibnamefont {Osborne}}, \bibinfo
  {author} {\bibfnamefont {I.}~\bibnamefont {Pi\ifmmode~\check{z}\else
  \v{z}\fi{}orn}}, \bibinfo {author} {\bibfnamefont {H.}~\bibnamefont
  {Verschelde}},\ and\ \bibinfo {author} {\bibfnamefont {F.}~\bibnamefont
  {Verstraete}},\ }\bibfield  {title} {\bibinfo {title} {Time-dependent
  variational principle for quantum lattices},\ }\href
  {https://doi.org/10.1103/PhysRevLett.107.070601} {\bibfield  {journal}
  {\bibinfo  {journal} {Phys. Rev. Lett.}\ }\textbf {\bibinfo {volume} {107}},\
  \bibinfo {pages} {070601} (\bibinfo {year} {2011})}\BibitemShut {NoStop}%
\bibitem [{\citenamefont {Haegeman}\ \emph {et~al.}(2016)\citenamefont
  {Haegeman}, \citenamefont {Lubich}, \citenamefont {Oseledets}, \citenamefont
  {Vandereycken},\ and\ \citenamefont {Verstraete}}]{Haegeman2016}%
  \BibitemOpen
  \bibfield  {author} {\bibinfo {author} {\bibfnamefont {J.}~\bibnamefont
  {Haegeman}}, \bibinfo {author} {\bibfnamefont {C.}~\bibnamefont {Lubich}},
  \bibinfo {author} {\bibfnamefont {I.}~\bibnamefont {Oseledets}}, \bibinfo
  {author} {\bibfnamefont {B.}~\bibnamefont {Vandereycken}},\ and\ \bibinfo
  {author} {\bibfnamefont {F.}~\bibnamefont {Verstraete}},\ }\bibfield  {title}
  {\bibinfo {title} {Unifying time evolution and optimization with matrix
  product states},\ }\href {https://doi.org/10.1103/PhysRevB.94.165116}
  {\bibfield  {journal} {\bibinfo  {journal} {Phys. Rev. B}\ }\textbf {\bibinfo
  {volume} {94}},\ \bibinfo {pages} {165116} (\bibinfo {year}
  {2016})}\BibitemShut {NoStop}%
\bibitem [{\citenamefont {Zauner-Stauber}\ \emph {et~al.}(2018)\citenamefont
  {Zauner-Stauber}, \citenamefont {Vanderstraeten}, \citenamefont {Fishman},
  \citenamefont {Verstraete},\ and\ \citenamefont {Haegeman}}]{Stauber2018}%
  \BibitemOpen
  \bibfield  {author} {\bibinfo {author} {\bibfnamefont {V.}~\bibnamefont
  {Zauner-Stauber}}, \bibinfo {author} {\bibfnamefont {L.}~\bibnamefont
  {Vanderstraeten}}, \bibinfo {author} {\bibfnamefont {M.~T.}\ \bibnamefont
  {Fishman}}, \bibinfo {author} {\bibfnamefont {F.}~\bibnamefont
  {Verstraete}},\ and\ \bibinfo {author} {\bibfnamefont {J.}~\bibnamefont
  {Haegeman}},\ }\bibfield  {title} {\bibinfo {title} {Variational optimization
  algorithms for uniform matrix product states},\ }\href
  {https://doi.org/10.1103/PhysRevB.97.045145} {\bibfield  {journal} {\bibinfo
  {journal} {Phys. Rev. B}\ }\textbf {\bibinfo {volume} {97}},\ \bibinfo
  {pages} {045145} (\bibinfo {year} {2018})}\BibitemShut {NoStop}%
\bibitem [{\citenamefont {Evertz}\ \emph {et~al.}(1993)\citenamefont {Evertz},
  \citenamefont {Lana},\ and\ \citenamefont {Marcu}}]{Evertz_1993}%
  \BibitemOpen
  \bibfield  {author} {\bibinfo {author} {\bibfnamefont {H.~G.}\ \bibnamefont
  {Evertz}}, \bibinfo {author} {\bibfnamefont {G.}~\bibnamefont {Lana}},\ and\
  \bibinfo {author} {\bibfnamefont {M.}~\bibnamefont {Marcu}},\ }\bibfield
  {title} {\bibinfo {title} {Cluster algorithm for vertex models},\ }\href
  {https://doi.org/10.1103/PhysRevLett.70.875} {\bibfield  {journal} {\bibinfo
  {journal} {Phys. Rev. Lett.}\ }\textbf {\bibinfo {volume} {70}},\ \bibinfo
  {pages} {875} (\bibinfo {year} {1993})}\BibitemShut {NoStop}%
\bibitem [{\citenamefont {Evertz}(2003)}]{Evertz_2003}%
  \BibitemOpen
  \bibfield  {author} {\bibinfo {author} {\bibfnamefont {H.~G.}\ \bibnamefont
  {Evertz}},\ }\bibfield  {title} {\bibinfo {title} {The loop algorithm},\
  }\href {https://doi.org/10.1080/0001873021000049195} {\bibfield  {journal}
  {\bibinfo  {journal} {Advances in Physics}\ }\textbf {\bibinfo {volume}
  {52}},\ \bibinfo {pages} {1–66} (\bibinfo {year} {2003})}\BibitemShut
  {NoStop}%
\bibitem [{\citenamefont {Yueh}\ and\ \citenamefont {Cheng}(2008)}]{Yueh2008}%
  \BibitemOpen
  \bibfield  {author} {\bibinfo {author} {\bibfnamefont {W.-C.}\ \bibnamefont
  {Yueh}}\ and\ \bibinfo {author} {\bibfnamefont {S.~S.}\ \bibnamefont
  {Cheng}},\ }\bibfield  {title} {\bibinfo {title} {Explicit eigenvalues and
  inverses of tridiagonal toeplitz matrices with four perturbed corners},\
  }\href {https://doi.org/10.1017/S1446181108000102} {\bibfield  {journal}
  {\bibinfo  {journal} {The ANZIAM Journal}\ }\textbf {\bibinfo {volume}
  {49}},\ \bibinfo {pages} {361–387} (\bibinfo {year} {2008})}\BibitemShut
  {NoStop}%
\bibitem [{\citenamefont {Abul-Magd}\ and\ \citenamefont
  {Abul-Magd}(2014)}]{Abul_Magd_2014}%
  \BibitemOpen
  \bibfield  {author} {\bibinfo {author} {\bibfnamefont {A.~A.}\ \bibnamefont
  {Abul-Magd}}\ and\ \bibinfo {author} {\bibfnamefont {A.~Y.}\ \bibnamefont
  {Abul-Magd}},\ }\bibfield  {title} {\bibinfo {title} {Unfolding of the
  spectrum for chaotic and mixed systems},\ }\href
  {https://doi.org/10.1016/j.physa.2013.11.012} {\bibfield  {journal} {\bibinfo
   {journal} {Physica A: Statistical Mechanics and its Applications}\ }\textbf
  {\bibinfo {volume} {396}},\ \bibinfo {pages} {185–194} (\bibinfo {year}
  {2014})}\BibitemShut {NoStop}%
\bibitem [{\citenamefont {Atas}\ \emph {et~al.}(2013)\citenamefont {Atas},
  \citenamefont {Bogomolny}, \citenamefont {Giraud},\ and\ \citenamefont
  {Roux}}]{Atas_2013}%
  \BibitemOpen
  \bibfield  {author} {\bibinfo {author} {\bibfnamefont {Y.~Y.}\ \bibnamefont
  {Atas}}, \bibinfo {author} {\bibfnamefont {E.}~\bibnamefont {Bogomolny}},
  \bibinfo {author} {\bibfnamefont {O.}~\bibnamefont {Giraud}},\ and\ \bibinfo
  {author} {\bibfnamefont {G.}~\bibnamefont {Roux}},\ }\bibfield  {title}
  {\bibinfo {title} {Distribution of the ratio of consecutive level spacings in
  random matrix ensembles},\ }\bibfield  {journal} {\bibinfo  {journal}
  {Physical Review Letters}\ }\textbf {\bibinfo {volume} {110}},\ \href
  {https://doi.org/10.1103/physrevlett.110.084101}
  {10.1103/physrevlett.110.084101} (\bibinfo {year} {2013})\BibitemShut
  {NoStop}%
\bibitem [{\citenamefont {Sachdev}(2011)}]{Sachdev_2011}%
  \BibitemOpen
  \bibfield  {author} {\bibinfo {author} {\bibfnamefont {S.}~\bibnamefont
  {Sachdev}},\ }\href@noop {} {\emph {\bibinfo {title} {Quantum Phase
  Transitions}}},\ \bibinfo {edition} {2nd}\ ed.\ (\bibinfo  {publisher}
  {Cambridge University Press},\ \bibinfo {year} {2011})\BibitemShut {NoStop}%
\bibitem [{\citenamefont {Fradkin}\ and\ \citenamefont
  {Susskind}(1978)}]{Fradkin_1978}%
  \BibitemOpen
  \bibfield  {author} {\bibinfo {author} {\bibfnamefont {E.}~\bibnamefont
  {Fradkin}}\ and\ \bibinfo {author} {\bibfnamefont {L.}~\bibnamefont
  {Susskind}},\ }\bibfield  {title} {\bibinfo {title} {Order and disorder in
  gauge systems and magnets},\ }\href
  {https://api.semanticscholar.org/CorpusID:18903916} {\bibfield  {journal}
  {\bibinfo  {journal} {Physical Review D}\ }\textbf {\bibinfo {volume} {17}},\
  \bibinfo {pages} {2637} (\bibinfo {year} {1978})}\BibitemShut {NoStop}%
\bibitem [{\citenamefont {Hasenfratz}\ \emph {et~al.}(1981)\citenamefont
  {Hasenfratz}, \citenamefont {Hasenfratz},\ and\ \citenamefont
  {Hasenfratz}}]{Hasenfratz_1981}%
  \BibitemOpen
  \bibfield  {author} {\bibinfo {author} {\bibfnamefont {A.}~\bibnamefont
  {Hasenfratz}}, \bibinfo {author} {\bibfnamefont {E.}~\bibnamefont
  {Hasenfratz}},\ and\ \bibinfo {author} {\bibfnamefont {P.}~\bibnamefont
  {Hasenfratz}},\ }\bibfield  {title} {\bibinfo {title} {Generalized roughening
  transition and its effect on the string tension},\ }\href
  {https://doi.org/https://doi.org/10.1016/0550-3213(81)90426-0} {\bibfield
  {journal} {\bibinfo  {journal} {Nuclear Physics B}\ }\textbf {\bibinfo
  {volume} {180}},\ \bibinfo {pages} {353} (\bibinfo {year}
  {1981})}\BibitemShut {NoStop}%
\bibitem [{\citenamefont {Sondhi}\ \emph {et~al.}(1997)\citenamefont {Sondhi},
  \citenamefont {Girvin}, \citenamefont {Carini},\ and\ \citenamefont
  {Shahar}}]{Sondhi1997}%
  \BibitemOpen
  \bibfield  {author} {\bibinfo {author} {\bibfnamefont {S.~L.}\ \bibnamefont
  {Sondhi}}, \bibinfo {author} {\bibfnamefont {S.~M.}\ \bibnamefont {Girvin}},
  \bibinfo {author} {\bibfnamefont {J.~P.}\ \bibnamefont {Carini}},\ and\
  \bibinfo {author} {\bibfnamefont {D.}~\bibnamefont {Shahar}},\ }\bibfield
  {title} {\bibinfo {title} {Continuous quantum phase transitions},\ }\href
  {https://doi.org/10.1103/RevModPhys.69.315} {\bibfield  {journal} {\bibinfo
  {journal} {Rev. Mod. Phys.}\ }\textbf {\bibinfo {volume} {69}},\ \bibinfo
  {pages} {315} (\bibinfo {year} {1997})}\BibitemShut {NoStop}%
\bibitem [{\citenamefont {Hasenbusch}(2005)}]{Hasenbusch2005}%
  \BibitemOpen
  \bibfield  {author} {\bibinfo {author} {\bibfnamefont {M.}~\bibnamefont
  {Hasenbusch}},\ }\bibfield  {title} {\bibinfo {title} {{The two-dimensional
  XY model at the transition temperature: a high-precision Monte Carlo
  study}},\ }\href {https://doi.org/10.1088/0305-4470/38/26/003} {\bibfield
  {journal} {\bibinfo  {journal} {Journal of Physics A: Mathematical and
  General}\ }\textbf {\bibinfo {volume} {38}},\ \bibinfo {pages} {5869}
  (\bibinfo {year} {2005})}\BibitemShut {NoStop}%
\bibitem [{\citenamefont {Cuesta}\ and\ \citenamefont
  {Sánchez}(2004)}]{Cuesta_2004}%
  \BibitemOpen
  \bibfield  {author} {\bibinfo {author} {\bibfnamefont {J.~A.}\ \bibnamefont
  {Cuesta}}\ and\ \bibinfo {author} {\bibfnamefont {A.}~\bibnamefont
  {Sánchez}},\ }\bibfield  {title} {\bibinfo {title} {General non-existence
  theorem for phase transitions in one-dimensional systems with short range
  interactions, and physical examples of such transitions},\ }\href
  {https://doi.org/10.1023/b:joss.0000022373.63640.4e} {\bibfield  {journal}
  {\bibinfo  {journal} {Journal of Statistical Physics}\ }\textbf {\bibinfo
  {volume} {115}},\ \bibinfo {pages} {869–893} (\bibinfo {year}
  {2004})}\BibitemShut {NoStop}%
\bibitem [{\citenamefont {{Zeiher}}\ \emph {et~al.}(2017)\citenamefont
  {{Zeiher}}, \citenamefont {{Choi}}, \citenamefont {{Rubio-Abadal}},
  \citenamefont {{Pohl}}, \citenamefont {{van Bijnen}}, \citenamefont
  {{Bloch}},\ and\ \citenamefont {{Gross}}}]{Zeiher_2017}%
  \BibitemOpen
  \bibfield  {author} {\bibinfo {author} {\bibfnamefont {J.}~\bibnamefont
  {{Zeiher}}}, \bibinfo {author} {\bibfnamefont {J.-y.}\ \bibnamefont
  {{Choi}}}, \bibinfo {author} {\bibfnamefont {A.}~\bibnamefont
  {{Rubio-Abadal}}}, \bibinfo {author} {\bibfnamefont {T.}~\bibnamefont
  {{Pohl}}}, \bibinfo {author} {\bibfnamefont {R.}~\bibnamefont {{van
  Bijnen}}}, \bibinfo {author} {\bibfnamefont {I.}~\bibnamefont {{Bloch}}},\
  and\ \bibinfo {author} {\bibfnamefont {C.}~\bibnamefont {{Gross}}},\
  }\bibfield  {title} {\bibinfo {title} {{Coherent Many-Body Spin Dynamics in a
  Long-Range Interacting Ising Chain}},\ }\href
  {https://doi.org/10.1103/PhysRevX.7.041063} {\bibfield  {journal} {\bibinfo
  {journal} {Physical Review X}\ }\textbf {\bibinfo {volume} {7}},\ \bibinfo
  {eid} {041063} (\bibinfo {year} {2017})},\ \Eprint
  {https://arxiv.org/abs/1705.08372} {arXiv:1705.08372 [physics.atom-ph]}
  \BibitemShut {NoStop}%
\bibitem [{\citenamefont {Bernien}\ \emph
  {et~al.}(2017{\natexlab{b}})\citenamefont {Bernien}, \citenamefont
  {Schwartz}, \citenamefont {Keesling}, \citenamefont {Levine}, \citenamefont
  {Omran}, \citenamefont {Pichler}, \citenamefont {Choi}, \citenamefont
  {Zibrov}, \citenamefont {Endres}, \citenamefont {Greiner}, \citenamefont
  {Vuleti{\'{c}}},\ and\ \citenamefont {Lukin}}]{Bernien2017}%
  \BibitemOpen
  \bibfield  {author} {\bibinfo {author} {\bibfnamefont {H.}~\bibnamefont
  {Bernien}}, \bibinfo {author} {\bibfnamefont {S.}~\bibnamefont {Schwartz}},
  \bibinfo {author} {\bibfnamefont {A.}~\bibnamefont {Keesling}}, \bibinfo
  {author} {\bibfnamefont {H.}~\bibnamefont {Levine}}, \bibinfo {author}
  {\bibfnamefont {A.}~\bibnamefont {Omran}}, \bibinfo {author} {\bibfnamefont
  {H.}~\bibnamefont {Pichler}}, \bibinfo {author} {\bibfnamefont
  {S.}~\bibnamefont {Choi}}, \bibinfo {author} {\bibfnamefont {A.~S.}\
  \bibnamefont {Zibrov}}, \bibinfo {author} {\bibfnamefont {M.}~\bibnamefont
  {Endres}}, \bibinfo {author} {\bibfnamefont {M.}~\bibnamefont {Greiner}},
  \bibinfo {author} {\bibfnamefont {V.}~\bibnamefont {Vuleti{\'{c}}}},\ and\
  \bibinfo {author} {\bibfnamefont {M.~D.}\ \bibnamefont {Lukin}},\ }\bibfield
  {title} {\bibinfo {title} {Probing many-body dynamics on a 51-atom quantum
  simulator},\ }\href {https://doi.org/10.1038/nature24622} {\bibfield
  {journal} {\bibinfo  {journal} {Nature}\ }\textbf {\bibinfo {volume} {551}},\
  \bibinfo {pages} {579} (\bibinfo {year} {2017}{\natexlab{b}})}\BibitemShut
  {NoStop}%
\bibitem [{\citenamefont {Scholl}\ \emph {et~al.}(2021)\citenamefont {Scholl},
  \citenamefont {Schuler}, \citenamefont {Williams}, \citenamefont
  {Eberharter}, \citenamefont {Barredo}, \citenamefont {Schymik}, \citenamefont
  {Lienhard}, \citenamefont {Henry}, \citenamefont {Lang}, \citenamefont
  {Lahaye}, \citenamefont {Läuchli},\ and\ \citenamefont
  {Browaeys}}]{Scholl2021}%
  \BibitemOpen
  \bibfield  {author} {\bibinfo {author} {\bibfnamefont {P.}~\bibnamefont
  {Scholl}}, \bibinfo {author} {\bibfnamefont {M.}~\bibnamefont {Schuler}},
  \bibinfo {author} {\bibfnamefont {H.~J.}\ \bibnamefont {Williams}}, \bibinfo
  {author} {\bibfnamefont {A.~A.}\ \bibnamefont {Eberharter}}, \bibinfo
  {author} {\bibfnamefont {D.}~\bibnamefont {Barredo}}, \bibinfo {author}
  {\bibfnamefont {K.-N.}\ \bibnamefont {Schymik}}, \bibinfo {author}
  {\bibfnamefont {V.}~\bibnamefont {Lienhard}}, \bibinfo {author}
  {\bibfnamefont {L.-P.}\ \bibnamefont {Henry}}, \bibinfo {author}
  {\bibfnamefont {T.~C.}\ \bibnamefont {Lang}}, \bibinfo {author}
  {\bibfnamefont {T.}~\bibnamefont {Lahaye}}, \bibinfo {author} {\bibfnamefont
  {A.~M.}\ \bibnamefont {Läuchli}},\ and\ \bibinfo {author} {\bibfnamefont
  {A.}~\bibnamefont {Browaeys}},\ }\bibfield  {title} {\bibinfo {title}
  {Quantum simulation of 2d antiferromagnets with hundreds of rydberg atoms},\
  }\href {https://doi.org/10.1038/s41586-021-03585-1} {\bibfield  {journal}
  {\bibinfo  {journal} {Nature}\ }\textbf {\bibinfo {volume} {595}},\ \bibinfo
  {pages} {233–238} (\bibinfo {year} {2021})}\BibitemShut {NoStop}%
\bibitem [{\citenamefont {Coleman}(1977)}]{Coleman_1977}%
  \BibitemOpen
  \bibfield  {author} {\bibinfo {author} {\bibfnamefont {S.}~\bibnamefont
  {Coleman}},\ }\bibfield  {title} {\bibinfo {title} {Fate of the false vacuum:
  Semiclassical theory},\ }\href {https://doi.org/10.1103/PhysRevD.15.2929}
  {\bibfield  {journal} {\bibinfo  {journal} {Phys. Rev. D}\ }\textbf {\bibinfo
  {volume} {15}},\ \bibinfo {pages} {2929} (\bibinfo {year}
  {1977})}\BibitemShut {NoStop}%
\bibitem [{\citenamefont {Lagnese}\ \emph {et~al.}(2021)\citenamefont
  {Lagnese}, \citenamefont {Surace}, \citenamefont {Kormos},\ and\
  \citenamefont {Calabrese}}]{Lagnese2021}%
  \BibitemOpen
  \bibfield  {author} {\bibinfo {author} {\bibfnamefont {G.}~\bibnamefont
  {Lagnese}}, \bibinfo {author} {\bibfnamefont {F.~M.}\ \bibnamefont {Surace}},
  \bibinfo {author} {\bibfnamefont {M.}~\bibnamefont {Kormos}},\ and\ \bibinfo
  {author} {\bibfnamefont {P.}~\bibnamefont {Calabrese}},\ }\bibfield  {title}
  {\bibinfo {title} {False vacuum decay in quantum spin chains},\ }\href
  {https://doi.org/10.1103/PhysRevB.104.L201106} {\bibfield  {journal}
  {\bibinfo  {journal} {Phys. Rev. B}\ }\textbf {\bibinfo {volume} {104}},\
  \bibinfo {pages} {L201106} (\bibinfo {year} {2021})}\BibitemShut {NoStop}%
\bibitem [{\citenamefont {Milsted}\ \emph {et~al.}(2022)\citenamefont
  {Milsted}, \citenamefont {Liu}, \citenamefont {Preskill},\ and\ \citenamefont
  {Vidal}}]{Milsted2022}%
  \BibitemOpen
  \bibfield  {author} {\bibinfo {author} {\bibfnamefont {A.}~\bibnamefont
  {Milsted}}, \bibinfo {author} {\bibfnamefont {J.}~\bibnamefont {Liu}},
  \bibinfo {author} {\bibfnamefont {J.}~\bibnamefont {Preskill}},\ and\
  \bibinfo {author} {\bibfnamefont {G.}~\bibnamefont {Vidal}},\ }\bibfield
  {title} {\bibinfo {title} {Collisions of false-vacuum bubble walls in a
  quantum spin chain},\ }\href {https://doi.org/10.1103/PRXQuantum.3.020316}
  {\bibfield  {journal} {\bibinfo  {journal} {PRX Quantum}\ }\textbf {\bibinfo
  {volume} {3}},\ \bibinfo {pages} {020316} (\bibinfo {year}
  {2022})}\BibitemShut {NoStop}%
\bibitem [{\citenamefont {Zenesini}\ \emph {et~al.}(2024)\citenamefont
  {Zenesini}, \citenamefont {Berti}, \citenamefont {Cominotti}, \citenamefont
  {Rogora}, \citenamefont {Moss}, \citenamefont {Billam}, \citenamefont
  {Carusotto}, \citenamefont {Lamporesi}, \citenamefont {Recati},\ and\
  \citenamefont {Ferrari}}]{Zenesini2024}%
  \BibitemOpen
  \bibfield  {author} {\bibinfo {author} {\bibfnamefont {A.}~\bibnamefont
  {Zenesini}}, \bibinfo {author} {\bibfnamefont {A.}~\bibnamefont {Berti}},
  \bibinfo {author} {\bibfnamefont {R.}~\bibnamefont {Cominotti}}, \bibinfo
  {author} {\bibfnamefont {C.}~\bibnamefont {Rogora}}, \bibinfo {author}
  {\bibfnamefont {I.~G.}\ \bibnamefont {Moss}}, \bibinfo {author}
  {\bibfnamefont {T.~P.}\ \bibnamefont {Billam}}, \bibinfo {author}
  {\bibfnamefont {I.}~\bibnamefont {Carusotto}}, \bibinfo {author}
  {\bibfnamefont {G.}~\bibnamefont {Lamporesi}}, \bibinfo {author}
  {\bibfnamefont {A.}~\bibnamefont {Recati}},\ and\ \bibinfo {author}
  {\bibfnamefont {G.}~\bibnamefont {Ferrari}},\ }\bibfield  {title} {\bibinfo
  {title} {False vacuum decay via bubble formation in ferromagnetic
  superfluids},\ }\href {https://doi.org/10.1038/s41567-023-02345-4} {\bibfield
   {journal} {\bibinfo  {journal} {Nature Physics}\ }\textbf {\bibinfo {volume}
  {20}},\ \bibinfo {pages} {558} (\bibinfo {year} {2024})}\BibitemShut
  {NoStop}%
\bibitem [{\citenamefont {Tausendpfund}\ \emph {et~al.}(2024)\citenamefont
  {Tausendpfund}, \citenamefont {Rizzi}, \citenamefont {Krinitsin},\ and\
  \citenamefont {Schmitt}}]{Tausendpfund2024}%
  \BibitemOpen
  \bibfield  {author} {\bibinfo {author} {\bibfnamefont {N.}~\bibnamefont
  {Tausendpfund}}, \bibinfo {author} {\bibfnamefont {M.}~\bibnamefont {Rizzi}},
  \bibinfo {author} {\bibfnamefont {W.}~\bibnamefont {Krinitsin}},\ and\
  \bibinfo {author} {\bibfnamefont {M.}~\bibnamefont {Schmitt}},\ }\href
  {https://doi.org/10.5281/zenodo.14421855} {\bibinfo {title} {{TTN -- A tree
  tensor network library for calculating groundstates and solving time
  evolution }}} (\bibinfo {year} {2024}),\ \bibinfo {note} {\textit{available
  on Zenodo.}}\BibitemShut {Stop}%
\bibitem [{\citenamefont {Fishman}\ \emph {et~al.}(2022)\citenamefont
  {Fishman}, \citenamefont {White},\ and\ \citenamefont
  {Stoudenmire}}]{Fishmann2022}%
  \BibitemOpen
  \bibfield  {author} {\bibinfo {author} {\bibfnamefont {M.}~\bibnamefont
  {Fishman}}, \bibinfo {author} {\bibfnamefont {S.~R.}\ \bibnamefont {White}},\
  and\ \bibinfo {author} {\bibfnamefont {E.~M.}\ \bibnamefont {Stoudenmire}},\
  }\bibfield  {title} {\bibinfo {title} {{The ITensor Software Library for
  Tensor Network Calculations}},\ }\href
  {https://doi.org/10.21468/SciPostPhysCodeb.4} {\bibfield  {journal} {\bibinfo
   {journal} {SciPost Phys. Codebases}\ ,\ \bibinfo {pages} {4}} (\bibinfo
  {year} {2022})}\BibitemShut {NoStop}%
\bibitem [{\citenamefont {Bauer}\ \emph {et~al.}(2011)\citenamefont {Bauer},
  \citenamefont {Carr}, \citenamefont {Evertz}, \citenamefont {Feiguin},
  \citenamefont {Freire}, \citenamefont {Fuchs}, \citenamefont {Gamper},
  \citenamefont {Gukelberger}, \citenamefont {Gull}, \citenamefont {Guertler},
  \citenamefont {Hehn}, \citenamefont {Igarashi}, \citenamefont {Isakov},
  \citenamefont {Koop}, \citenamefont {Ma}, \citenamefont {Mates},
  \citenamefont {Matsuo}, \citenamefont {Parcollet}, \citenamefont
  {Pawłowski}, \citenamefont {Picon}, \citenamefont {Pollet}, \citenamefont
  {Santos}, \citenamefont {Scarola}, \citenamefont {Schollwöck}, \citenamefont
  {Silva}, \citenamefont {Surer}, \citenamefont {Todo}, \citenamefont {Trebst},
  \citenamefont {Troyer}, \citenamefont {Wall}, \citenamefont {Werner},\ and\
  \citenamefont {Wessel}}]{Bauer_2011}%
  \BibitemOpen
  \bibfield  {author} {\bibinfo {author} {\bibfnamefont {B.}~\bibnamefont
  {Bauer}}, \bibinfo {author} {\bibfnamefont {L.~D.}\ \bibnamefont {Carr}},
  \bibinfo {author} {\bibfnamefont {H.~G.}\ \bibnamefont {Evertz}}, \bibinfo
  {author} {\bibfnamefont {A.}~\bibnamefont {Feiguin}}, \bibinfo {author}
  {\bibfnamefont {J.}~\bibnamefont {Freire}}, \bibinfo {author} {\bibfnamefont
  {S.}~\bibnamefont {Fuchs}}, \bibinfo {author} {\bibfnamefont
  {L.}~\bibnamefont {Gamper}}, \bibinfo {author} {\bibfnamefont
  {J.}~\bibnamefont {Gukelberger}}, \bibinfo {author} {\bibfnamefont
  {E.}~\bibnamefont {Gull}}, \bibinfo {author} {\bibfnamefont {S.}~\bibnamefont
  {Guertler}}, \bibinfo {author} {\bibfnamefont {A.}~\bibnamefont {Hehn}},
  \bibinfo {author} {\bibfnamefont {R.}~\bibnamefont {Igarashi}}, \bibinfo
  {author} {\bibfnamefont {S.~V.}\ \bibnamefont {Isakov}}, \bibinfo {author}
  {\bibfnamefont {D.}~\bibnamefont {Koop}}, \bibinfo {author} {\bibfnamefont
  {P.~N.}\ \bibnamefont {Ma}}, \bibinfo {author} {\bibfnamefont
  {P.}~\bibnamefont {Mates}}, \bibinfo {author} {\bibfnamefont
  {H.}~\bibnamefont {Matsuo}}, \bibinfo {author} {\bibfnamefont
  {O.}~\bibnamefont {Parcollet}}, \bibinfo {author} {\bibfnamefont
  {G.}~\bibnamefont {Pawłowski}}, \bibinfo {author} {\bibfnamefont {J.~D.}\
  \bibnamefont {Picon}}, \bibinfo {author} {\bibfnamefont {L.}~\bibnamefont
  {Pollet}}, \bibinfo {author} {\bibfnamefont {E.}~\bibnamefont {Santos}},
  \bibinfo {author} {\bibfnamefont {V.~W.}\ \bibnamefont {Scarola}}, \bibinfo
  {author} {\bibfnamefont {U.}~\bibnamefont {Schollwöck}}, \bibinfo {author}
  {\bibfnamefont {C.}~\bibnamefont {Silva}}, \bibinfo {author} {\bibfnamefont
  {B.}~\bibnamefont {Surer}}, \bibinfo {author} {\bibfnamefont
  {S.}~\bibnamefont {Todo}}, \bibinfo {author} {\bibfnamefont {S.}~\bibnamefont
  {Trebst}}, \bibinfo {author} {\bibfnamefont {M.}~\bibnamefont {Troyer}},
  \bibinfo {author} {\bibfnamefont {M.~L.}\ \bibnamefont {Wall}}, \bibinfo
  {author} {\bibfnamefont {P.}~\bibnamefont {Werner}},\ and\ \bibinfo {author}
  {\bibfnamefont {S.}~\bibnamefont {Wessel}},\ }\bibfield  {title} {\bibinfo
  {title} {{The ALPS project release 2.0: open source software for strongly
  correlated systems}},\ }\href
  {https://doi.org/10.1088/1742-5468/2011/05/P05001} {\bibfield  {journal}
  {\bibinfo  {journal} {Journal of Statistical Mechanics: Theory and
  Experiment}\ }\textbf {\bibinfo {volume} {2011}},\ \bibinfo {pages} {P05001}
  (\bibinfo {year} {2011})}\BibitemShut {NoStop}%
\bibitem [{\citenamefont {{J\"{u}lich Supercomputing Centre}}(2021)}]{JUWELS}%
  \BibitemOpen
  \bibfield  {author} {\bibinfo {author} {\bibnamefont {{J\"{u}lich
  Supercomputing Centre}}},\ }\bibfield  {title} {\bibinfo {title} {{JUWELS
  Cluster and Booster: Exascale Pathfinder with Modular Supercomputing
  Architecture at JSC}},\ }\href {https://doi.org/10.17815/jlsrf-7-183}
  {\bibfield  {journal} {\bibinfo  {journal} {Journal of large-scale research
  facilities}\ }\textbf {\bibinfo {volume} {7}},\ \bibinfo {pages} {A183}
  (\bibinfo {year} {2021})}\BibitemShut {NoStop}%
\bibitem [{\citenamefont {{J{\"u}lich Supercomputing
  Centre}}(2021)}]{JURECA2021}%
  \BibitemOpen
  \bibfield  {author} {\bibinfo {author} {\bibnamefont {{J{\"u}lich
  Supercomputing Centre}}},\ }\bibfield  {title} {\bibinfo {title} {{JURECA:
  Data Centric and Booster Modules implementing the Modular Supercomputing
  Architecture at JSC}},\ }\href {https://doi.org/10.17815/jlsrf-7-182}
  {\bibfield  {journal} {\bibinfo  {journal} {Journal of large-scale research
  facilities}\ }\textbf {\bibinfo {volume} {7}},\ \bibinfo {pages} {A182}
  (\bibinfo {year} {2021})}\BibitemShut {NoStop}%
\bibitem [{\citenamefont {Krinitsin}\ \emph
  {et~al.}(2025{\natexlab{b}})\citenamefont {Krinitsin}, \citenamefont
  {Tausendpfund}, \citenamefont {Rizzi}, \citenamefont {Heyl},\ and\
  \citenamefont {Schmitt}}]{krinitsin_2025}%
  \BibitemOpen
  \bibfield  {author} {\bibinfo {author} {\bibfnamefont {W.}~\bibnamefont
  {Krinitsin}}, \bibinfo {author} {\bibfnamefont {N.}~\bibnamefont
  {Tausendpfund}}, \bibinfo {author} {\bibfnamefont {M.}~\bibnamefont {Rizzi}},
  \bibinfo {author} {\bibfnamefont {M.}~\bibnamefont {Heyl}},\ and\ \bibinfo
  {author} {\bibfnamefont {M.}~\bibnamefont {Schmitt}},\ }\bibfield  {title}
  {\bibinfo {title} {Data and code associated to the paper ``{Roughening
  dynamics of interfaces in two-dimensional quantum matter}''},\ }\href
  {https://doi.org/10.5281/zenodo.14705609} {10.5281/zenodo.14705609} (\bibinfo
  {year} {2025}{\natexlab{b}})\BibitemShut {NoStop}%
\bibitem [{\citenamefont {Hearth}(2024)}]{Hearth2024}%
  \BibitemOpen
  \bibfield  {author} {\bibinfo {author} {\bibfnamefont {S.}~\bibnamefont
  {Hearth}},\ }\href {https://github.com/Renmusxd/IsingMonteCarlo} {\bibinfo
  {title} {{IsingMonteCarlo}}} (\bibinfo {year} {2024})\BibitemShut {NoStop}%
\end{thebibliography}%
